\documentclass[aps,twocolumn,floats,prc]{revtex4}
\usepackage{graphics,graphicx,epsfig}
\usepackage{amssymb,color}
\usepackage{epsf,epstopdf,wrapfig}
\pdfoutput=1

\newcommand{\abs}[1]{|#1|}
\newcommand{\av}[1]{\langle{#1}\rangle{}}
\newcommand{\beq}{\begin{equation}}
\newcommand{\eeq}{\end{equation}}
\newcommand{\beqn}{\begin{eqnarray}}
\newcommand{\eeqn}{\end{eqnarray}}
\newcommand {\e}[1]{\mathrm{~#1}}    
		
\begin{document}

\title{Information transmission in genetic regulatory networks: a review}

\author{Ga\v{s}per Tka\v{c}ik$^{1}$\footnote{gtkacik@ist.ac.at}  }

\author{Aleksandra M. Walczak$^{2}$\footnote{awalczak@lpt.ens.fr}  }

\affiliation{\mbox{${}^1$Institute of Science and Technology Austria, Am Campus 1, A-3400 Klosterneuburg, Austria} \\ ${}^{2}$CNRS-Laboratoire de Physique Th\'eorique de l'\'Ecole Normale Sup\'erieure, \mbox{24 rue Lhomond, 75005 Paris, France}}

\date{\today}

\begin{abstract}
Genetic regulatory networks enable cells to respond to the changes in internal and external conditions by dynamically coordinating their gene expression profiles. 
Our ability to make quantitative measurements in these biochemical circuits has deepened our understanding of what kinds of computations genetic regulatory networks can perform and with what reliability. These advances have motivated researchers to look for connections between the architecture and function of genetic regulatory networks. Transmitting information between network's inputs and its outputs  has been proposed as one such possible measure of function, relevant in certain biological contexts. Here we summarize recent developments in the application of information theory to gene regulatory networks. We first review basic concepts in information theory necessary to understand recent work.  We then discuss the functional complexity of gene regulation which arrises from the molecular nature of the regulatory interactions. We end by reviewing some experiments supporting the view that genetic networks responsible for early development of multicellular organisms  might be maximizing transmitted ``positional'' information.
\end{abstract}

\maketitle
\tableofcontents
\section{Introduction}
In the classical view of genetics, the information necessary for the functioning of a given organism is encoded in its DNA  \cite{alberts, lewin}. Gene expression is a process by which this information is extracted from the DNA in order to synthesize proteins that carry out specific functions in the cell. For instance, actin and tubulin provide structural support, myosin can generate physical forces, kinases and phosphatases are instrumental in intracellular signaling pathways, substrate-specific enzymes drive the metabolic cycle, and, ultimately, gene expression machinery itself needs to be synthesized from its DNA blueprint.  According to the central dogma, information flows from DNA to proteins: first the genes on DNA are transcribed into mRNA, which is converted by the ribosomes  into amino acid sequences that fold into functioning proteins. It is quite obvious, however, that information must flow in the other direction as well, dictating under what conditions which proteins should be produced from their DNA blueprints. The best example of this are multicellular organisms: although all of their cells share the same genomic DNA, they do not all express the same proteins, and it is this selective gene expression that allows the cells to specialize into different phenotypes, build up a range of tissues, and fulfill specific organismal functions. 

All cellular processes which control the expression of proteins are collectively called \emph{gene regulation}.
Gene regulation can occur at essentially every step of extracting the information from the DNA: at the level of DNA packing and epigenetic modifications, at transcription initiation, translation, through modifications of mRNAs, or through post-translational modifications of amino-acid sequences \cite{Wagner}. These processes  are mainly effected by special proteins with regulatory function, among which we single out as a prominent example \emph{transcription factors} that can modify the transcription activity at their target genes. At any moment, the state of a living organism is thus not described by its genome alone, but also by the set of (regulatory) genes that the organism actually expresses and the concentration levels of the corresponding gene products. 

The possible phenotypic states of a cell correspond to distinct gene expression patterns. In this view, DNA and the associated regulation machinery give rise to a finite yet large number of \emph{possible} cellular ``outcome'' states, while the \emph{actual} state is selected from this possible range both by current internal and environmental conditions. Despite recent experimental and theoretical progress in characterizing molecular properties of various regulatory subunits and specific molecular pathways, we do not fully understand how these elements come together to form a functioning system and how precisely they fit into the conceptual picture outlined above. Classic genetic experiments on model systems, as well as bioinformatics in conjunction with high-throughput assays, have started to fill out the map of regulatory interactions in the cell, i.e. which transcription factor proteins regulate which genes, which pathways are activated under given conditions, and what is the role of non-transcriptional regulatory mechanisms. What we are learning in terms of detail, however, is opening up new questions on the systems level: in trying to understand the experimentally reconstructed regulatory networks, we find them statistically far from random, but also far from how human engineers would go about solving the problems that cells are presumably trying to solve. Our difficulty in understanding and reverse-engineering these networks gives rise to several important questions: Why do regulatory networks have the observed architectures?  Are we correctly and quantitatively understanding the functions that they perform? Can we look for factors that discriminate the networks that exist in nature as opposed to the ones that do not? Are existing networks simply artifacts of evolutionary history, or are there features that discriminate them from the non-existent but in principle possible architectures? Are observed forms of gene regulation all necessary in different contexts or are they simply redundant? Can we go beyond mere characterizations of regulatory networks and instead identify physical principles that govern the observed network behaviors?

A number of groups have recently explored different physical principles that could influence the parameter regimes and modes of regulation in living organisms \cite{franhakimsiggia,francsiggia,massimo,tostevintenwoldehoward,ulihwapnas09, ggpnas,pankaj}. Such approaches usually require one to choose a measure of network function. Among options being considered are minimization of biochemical noise \cite{tostevintenwoldehoward, saundershoward}, optimization of losses in case of unknown enviromental signals \cite{massimo}, maximization of positional information \cite{ggpnas, francoissiggia}, or optimization of resources \cite{fromulispaper}. Some of these strategies have also been considered in the presence of evolutionary forces \cite{ulihwapnas09,francoissiggia}. Here, we review the work that has focussed on optimizing information transmission in gene regulatory networks. 

Assuming that information transmission is a viable measure of network function, we can explore and compare various network architectures and modes of gene regulation. We note upfront that the assumption we are making is a strong one and in general gene regulatory networks need not be optimized at all; nevertheless, we claim that {\bf (i)} in certain biological contexts this assumption might be close to valid; {\bf (ii)} it will enable us to make some analytic progress; {\bf (iii)}  even if not fully correct, such assumption allows us to make experimentally testable predictions. We argue these points in detail in Section \ref{Sec_justifyinfo}. We start by introducing a mathematical framework in which gene regulation can be described (Section \ref{Sec_molmodels}). We then formalize the concept of information (Section \ref{info_why}, Section \ref{info}), proceed to review optimal networks in the limit of small noise (Section \ref{SNA}, Section \ref{ONA}) and beyond this limit (Section \ref{beyondSNA}). Lastly we discuss information transmission in the presence of time-dependent signals (Section \ref{timedep}).

This review is primarily aimed at a physics readership, but we hope it can be enjoyed by anyone with an interest in the interface between information theory and gene regulation. We methodologically introduce both topics, neither of which is typically discussed in the traditional physics curriculum. All results presented in this review have been published elsewhere; parts of this review follow the exposition of Ref~\cite{oist} which discusses the links between statistical physics and biological networks in greater detail. To keep bibliography manageable, we decided to reference solely standard textbooks, papers that directly discuss signal transmission in biological networks, and experimental papers that we provide as examples in this manuscript.  We do not provide extensive referencing for large and relevant fields discussing noise in gene expression, statistical and dynamical properties of regulatory networks in general, or papers providing biological detail on early embryonic development;  interested readers should consult Ref~\cite{enc} and references therein.

\section{Functional aspects of gene regulation}\label{Sec_molmodels}

The expression of genes in cells is controlled mainly by binding and unbinding of regulatory proteins, called \emph{transcription factors} (TFs), to specific short DNA sequences, called  \emph{binding sites} \cite{ptashne}. These regulatory proteins can act either as \emph{activators}, which means they increase the rate of expression of the genes, or as \emph{repressors} that decrease the rate of expression of the regulated genes. The genetic sequence of the DNA is  \emph{transcribed} into mRNA by a holoenzyme called RNA polymerase. Activators often act by recruiting the polymerase, whereas repressors often act by sterically blocking the polymerase from binding. Ribosomes  \emph{translate} mRNA strands into proteins. TFs can cross- and self-regulate, opening up a possibility of feedback regulation. They are usually present in nuclei in small, nanomolar range concentrations (for a nucleus with several $\e{\mu m}$ radius, these concentrations correspond to several hundred to thousands of TF molecules per nucleus). The timescales of such regulation span a wide range, from minutes to hours. 

Generally, the expression of genes can be regulated at all levels, from DNA looping to post-translational modification of proteins. Often, many co-factors and enzymes are involved, and the process can be described in a molecularly detailed fashion. However, certain features can be abstracted and allow us to study generalized models of gene expression:
\begin{itemize}
\item Regulation functions, i.e. functions that map the concentrations of TFs into levels of regulated gene expression, in gene regulatory networks are {\bf nonlinear}. There are saturation effects, for example when a gene is fully activated. Nonlinearities in regulation also set the range of input concentrations in which a network is responsive. In addition to the simple nonlinearities induced by saturation effects, networks often contain positive or negative {\bf feedback loops} that can give rise to even more complicated behaviors.
\item  Gene regulation is a {\bf noisy} process. This is a consequence of the stochasticity in single molecular events at low concentrations of the relevant molecules, such as in reactions between TFs and binding sites (that can be present at copy numbers of only one or two in the whole genome). The nanomolar concentrations of TFs in the cell mean that the precise timing when a TF finds and binds a regulatory site on the DNA is a random variable; this randomness results in stochastic gene activation. 
\item The processes involved in gene regulation happen on various time scales: the time on which the input fluctuates, the protein decay time, the gene expression state fluctuation time, the time on which the external input signal changes. The networks are {\bf dynamical systems}, and their behaviors span the range from settling down to one of the possible stationary states, to generating intrinsic oscillations (as in, e.g., circadian clocks) or more complex combinations of checkpoint steady states and limit cycle oscillations (as in cell-cycle control).
\item The wiring in the network is {\bf specific}. Specificity is achieved by molecular mechanisms of recognition (TF--DNA interaction). One TF can regulate many genes by recognizing and binding multiple sites in the genome, and each gene can be regulated by several TFs.
\end{itemize} 

One can describe a gene regulatory circuit at various levels of detail. All of them attempt to capture most of the properties listed above, with different emphasis on the particular points (see Refs~\cite{enc, walczak_chapter} for more information). Here we will briefly review a few basic approaches that we are going to use later in this review, on a specific example of a single regulatory element.

\subsection{Gene regulatory elements: a mathematical primer}
\begin{figure}
\includegraphics[width = 0.6 \linewidth]{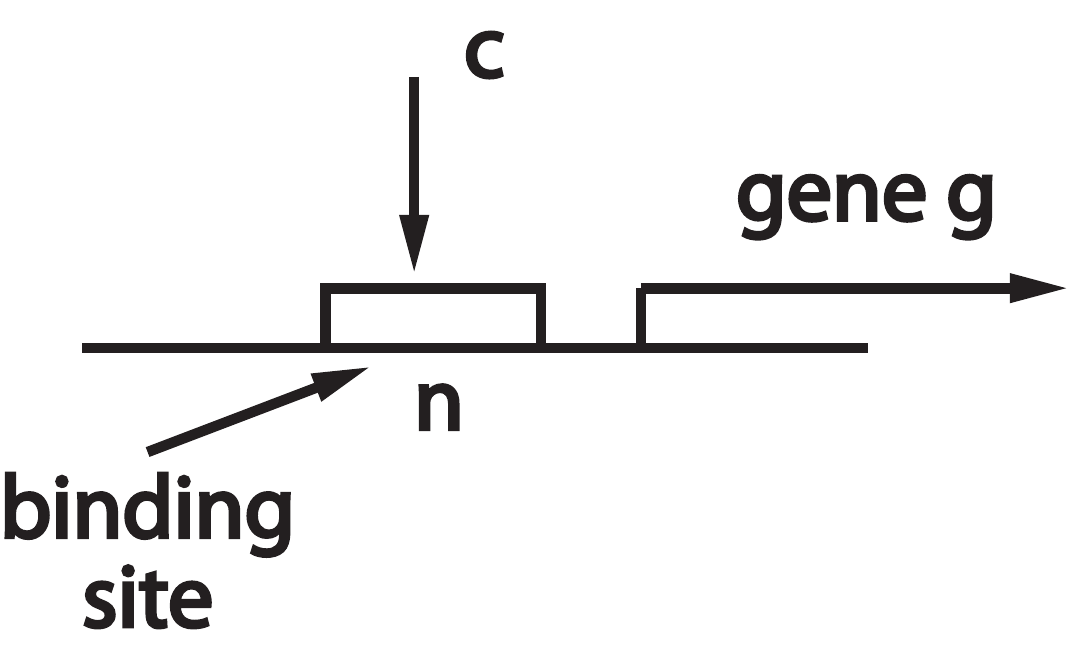}
\caption{The simplest regulatory graph, where an input transcription factor at concentration $c$ regulates the output expression level $g$ by binding to a binding site $n$, which can be empty or occupied. Since $c$ acts as an activator, an occupied site results in transcription and translation of $g$.}
\label{f-sscheme}
\end{figure}

Let transcription factors  be present at concentration $c$ in the cell. On the DNA, there is a single specific binding site that can be occupied or empty; we will denote this occupancy with $n(t)$. When the site is occupied, the regulated gene will get transcribed into mRNA, which is later translated into proteins whose count we denote by $g(t)$, at the combined rate that we denote by $R$. The proteins are degraded with the characteristic time $\tau$.  In this case, our TF thus acts as an activator, see Fig.~\ref{f-sscheme}. Here and afterwards we will refer to the transcription factor $c$ as an \emph{input}, and the regulated gene product $g$ as \emph{output}.

This model discards a lot of molecular complexity: there is no explicit treatment of diffusion of TFs, no non-specific binding,  no separate treatment of mRNA and protein, no chromatin opening / closing etc; in addition, we group many multi-stage molecular processes (such as TF binding, RNAP assembly, processive transcription etc) into single coarse-grained steps. Thus, our model is a gross (but tractable) oversimplification. As an illustration, let us formulate it in a few different mathematical frameworks.

In the limit of relatively large concentrations, we can treat concentrations $c$ and $g$ as continuous and describe this regulatory process by the set of differential equations for the means of the concentrations:
\begin{eqnarray}
\frac{dn}{dt} &=& k_+ c(t)(1-n) - k_-n	\label{occ}\\
\frac{dg}{dt}&=&-\frac{1}{\tau}g + R n. \label{prot}
\end{eqnarray}
Equation (\ref{occ}) is an equation for occupancy $n$, which is a number between 0 and 1. Nominally, the site can only be fully empty or occupied, but in this approximation, we treat it as a continuous variable that can be interpreted as a ``probability of the site being bound.'' $k_+c$ is the TF-concentration-dependent on-rate, and $k_-$ is the first-order off-rate. Often, it is assumed that there is a separation of time scales: the first equation for occupancy equilibrates much faster than $\tau$, meaning that the mean occupancy 
\begin{equation}
\bar{n}(t) = \frac{k_+ c(t)}{k_+c(t) + k_-}
\end{equation}
can be inserted into Eq (\ref{prot}) to get
\begin{equation}
\frac{dg}{dt}=-\frac{1}{\tau}g(t) + R\frac{k_+ c(t)}{k_+c(t) + k_-}. \label{prot1}
\end{equation}
In this simple case without feedback, the approach to the equilibrium at fixed $c$ is exponential with the rate $\tau$, and the steady state is simple: $\bar{g} = R\tau \bar{n}$. The effective production rate $R \bar{n}$ in Eq~(\ref{prot1}) is a function with a sigmoidal shape. We discuss in the next section how the particular sigmoidal regulation functions are connected to equilibrium statistical mechanics of this system, how noise can be added by an introduction of the Langevin force into Eq (\ref{prot}), and why the assumption of fast equilibration of $n$ strongly influences the noise. 

%% what is noise?

Suppose we wanted to capture the idea that the number of molecules in the system is discrete and that reactions between them are stochastic. In this case the object of our inquiry would be $P_n(g|t,c)$: the time-dependent joint probability of observing $g$ molecules of the resulting gene and the state of the binding site being $n=0,1$ (empty, occupied), given some concentration of the input $c$. One can marginalize this distribution over $n$ to get the evolution of probability of observing $g$ output molecules: $P(g|t,c)=\sum_{n=0,1}P_n(g|t,c)$. Writing down the master equation \cite{vanKampen,Gardner} and for simplicity suppressing the parameters $(c,t)$ on which all terms $P_n(g|t,c)$ are conditioned, we find:
\begin{eqnarray}
\frac{dP_0(g)}{dt} &=& \frac{g+1}{\tau}P_0(g+1)+ k_-P_1(g) - (k_+ c + \frac{g}{\tau})P_0(g) \nonumber\\
\frac{dP_1(g)}{dt} &=& \frac{g+1}{\tau}P_1(g+1) +RP_1(g-1)+k_+cP_0(g) -\nonumber\\
&-& (k_- + \frac{g}{\tau}+R)P_1(g); \label{mastereq}
\end{eqnarray}
the reader should recognize degradation-related terms (proportional to $1/\tau$), the protein production terms (prefixed with $R$ and present only in the case when the gene is on, i.e. $n=1$) and the switching terms of the promoter containing $k_+ c$ and $k_-$, which couple the $n=0$ to $n=1$ states.  In this simple case, the equilibrium distribution can be solved by zeroing out the left-hand side of Eqs~(\ref{mastereq}). This yields an infinite dimensional system in $g$ that can be truncated at some $g_{\rm max}\gg R\tau$; we would end up with a homogenous linear system that can be supplemented by a normalization condition $\sum_{n=0,1}\sum_{g=0}^{g_{\rm max}}P_n(g)=1$, which can be inverted and solved for steady state $P_n(g)$. More sophisticated methods are available when the number of genes grows and they are interacting \cite{wmw}. Note that in this example we treated $g$ as discrete, but $c$ is still a continuous input parameter (not a variable whose distributions we are also interested in). We can directly calculate the moments $\langle g^k\rangle= \sum_{g,n} g^k P_n(g)$ from the steady state master equations. If we define  $n=\sum_g P_1(g)$, we reproduce (with $k=1$) the equations for the averages $g(t)$, $n(t)$ in Eqs~(\ref{occ},\ref{prot}). 

One can also expand the master equation to second order. If we assume that the gene expression state changes on fast timescales compared to the change in the number of proteins, we obtain the Fokker-Planck equation for $P(g)=P_0(g)+P_1(g)$:
\begin{eqnarray}
\frac{dP(g)}{dt} &=&\frac{\partial}{\partial g} \left[(R\frac{k_+ c(t)}{k_+c(t) + k_-}-g/\tau) P(g)\right]\\ \nonumber
&&-\frac{1}{2}\frac{\partial^2}{\partial g^2} \left[(R\frac{k_+ c(t)}{k_+c(t) + k_-}+g/\tau) P(g)\right].\label{FP0}
\end{eqnarray}
Equation (\ref{prot1}) can be recovered again by calculating the mean of $g$ from the Fokker-Planck equations.  Both the master Eqs~(\ref{mastereq}) and Fokker-Planck Eq~(\ref{FP0}) allow us to calculate higher order moments apart from the mean; for instance, by computing the second moments, we can write down the fluctuation of the number of proteins around its mean, $\sigma_g^2(t,c) = \langle g^2\rangle - \langle g \rangle^2$. This is intrinsic noise, or stochasticity in gene expression due to the randomness and discrete nature of molecular interactions. 

If we assume \emph{a priori} that a gene regulatory process can be described well in terms of the dynamics for the mean values [as in Eq~(\ref{occ},\ref{prot})] plus a Gaussian fluctuation around the mean (ignoring higher order moments), there exists a systematic procedure for calculating the resulting noise variances, called the Langevin approximation. In this approximation one starts by writing down ordinary differential equations for the mean values, and adds an \emph{ad hoc} noise force, 
\begin{equation}
\frac{dg}{dt}=-\frac{1}{\tau}g(t) + R\frac{k_+ c(t)}{k_+c(t) + k_-} +\xi(t). \label{Lanprot1}
\end{equation}
The ``random force'' term $\xi$ takes a form that we need to assume based on physical intuition and more formal methods (e.g. the Fokker-Planck equation). In this case we can postulate that the noise magnitude $T$ depends on the state of the system, but that fluctuations are zero mean, $\langle \xi(t)\rangle=0$, random and uncorrelated in time, i.e. $\langle \xi(t)\xi(t')\rangle=2T(g)\delta(t-t')$  (braces denote averaging over many realizations of the noise time series). We'll discuss noise in gene expression in detail later, including a worked-out example using Langevin approximation. For a detailed derivation of various approximations in gene regulation see Ref~\cite{walczak_chapter}.

Finally, let us mention the numerical Gillespie algorithm \cite{Gillespie}. For this algorithm we start with enumerating all reactions $i$ and their rates $r_i$:
\begin{eqnarray}
r_1= k_{+\;} &:& c+n\rightarrow cn	 \nonumber\\
r_2= k_{-\;} &:& cn \rightarrow c + n \nonumber\\
r_3= R_{\;\;\; }&:& cn \rightarrow cn + g \nonumber \\
r_4=\tau^{-1}&:& g \rightarrow \oslash \label{cgillespie}
\end{eqnarray}
The state of the system is then initialized as a vector $(c,n,cn, g)$ of integer counts of molecular species (here $cn$ denotes a molecular complex of a $c$ molecule bound to the promoter; there can only be 0 or 1 $n$ and $cn$, and one can quickly check  that $n=1-cn$). Then the probability per unit time  of each of the 4 reactions is the product of the rate constant $r_i$ and the number of reactants properly normalized by the relevant volume.
The algorithm randomly draws the next reaction consistent with the probabilities per unit time, updates the state of the system and repeats. This algorithm is exact for well-mixed systems, but {\bf (i)} it can be slow in case there are fast and slow reactions in the system; {\bf (ii)} one needs to sample many simulation runs to accumulate the noise statistics; {\bf (iii)} it can become incorrect in biological systems where transport (e.g. diffusion) needs to be taken into account explicitly \cite{simapaper,vanzon,tkacik+al_09}.

From the presented example it is clear that the fully stochastic dynamical description can be relatively complicated even for a very simple system. To proceed and be able to connect to data, we will drop the time dependence and only focus on the steady state, while emphasizing the nonlinear and noisy nature of the system. Our assumption to only study the steady state will preclude us from discussing network phenomena that are intrinsically dynamic, e.g. the cell cycle or the circadian clock. But for many biologically realistic cases, such as in developmental biology, or in many experimental settings, such as measuring the gene response to  constant levels of inducer, the steady state approach is useful.
\subsection{Regulation by a single transcription factor}
In this section we will explore simple thermodynamical models of gene regulation, by studying how the concentration of a transcription factor relates to promotor occupancy and thus to the expression level of the regulated gene. A detailed discussion of the thermodynamic approach to gene regulation with worked out examples for various regulatory strategies can be found in Refs~\cite{BintuHwa1, BintuHwa2}.

In the previous section we saw that we can obtain the expression for the mean promoter occupancy directly from the master equation, assuming that the system is in equilibrium. Under this assumption we can ask for the equivalent statistical mechanics description which, as we shall see, can be easily generalized to larger systems. 

Suppose we have a site $n$ that can be occupied or empty. In case it is occupied, there is a binding energy $E$ favoring the occupied state, relative to the reference energy $0$ in the unbound state. But in order to occupy the state, one needs to remove one molecule of TF  from the solution. The chemical potential of TFs, or the free energy cost of removing a single molecule of TF from the solution, is $\mu=k_B T \log c$, where $c$ is the TF concentration measured in some dimensionless units of choice. 
In statistical physics we can calculate every equilibrium property of the system if we know how to compute the \emph{partition sum}, which is $Z=\sum_i e^{-\beta(E_i - \mu n_i)}$, where the sum is taken over all possible states of the system (in our case binding site empty and binding site occupied), $E_i$ is the energy of the system is the state $i$, and $n_i$ is the number of molecules in the system in the state $i$. 

In our case of a single binding site, the partition sum is taken over the empty ($n=0$) and occupied ($n=1$) states:
\begin{equation}
Z=e^{-\beta (E-\mu)}+1,
\end{equation}
where $\beta=1/(k_BT)$, $T$ is the temperature in Kelvin and $k_B$ is the Boltzmann constant.  The probability that the site is occupied is then
\begin{equation}
P(n=1) = \frac{1}{Z}e^{-\beta (E-\mu)}.
\end{equation}
Inserting the definition of $\mu$, we get
\begin{equation}
P(n=1) = \frac{c}{c+K_d}, \label{occaux}
\end{equation}
where we write $K_d=\exp(\beta E)$. But $\bar{n} = 1\cdot P(n=1) + 0\cdot P(n=0) = P(n=1)$, so by comparing with Eq~(\ref{occ}) we can make the identification
\begin{equation}
K_d = e^{\beta E}= \frac{k_-}{k_+}, \label{db}
\end{equation}
which connects our statistical mechanics and dynamical pictures. Note that $k_-$ is measured in units of inverse time, $\e{s}^{-1}$, $k_+$ is measured in units of $s^{-1}\times [\mathrm{conc}]^{-1}$ (but by convention we here measure concentration in dimensionless units, as in $\mu = k_BT\log c$), so $K_d$ has units of concentration.

Suppose we make the model somewhat more complicated: let us have two binding sites, which together will constitute a system with 4 possible states of occupancy: both sites empty, either one occupied, and both occupied, which we will write compactly as $(00,01,10,11)$. Let us also assume that there is cooperativity in the system -- if both sites are occupied, then there will be an additional favorable energetic contribution of $\epsilon$ to the total energy of the state $(11)$. Finally, when promoters can have multiple internal states, we need to decide which state is the ``active'' state, when the gene is being transcribed\footnote{In general, each internal promoter state could have its own transcription rate, but often one state is picked as having the maximal transcription rate, and the other states represent the gene being ``off'' or expressing at some small, ``leaky'' rate of gene expression.}; here we pick the state $(11)$ as the active state.

\begin{figure}
\includegraphics[width = 0.8 \linewidth]{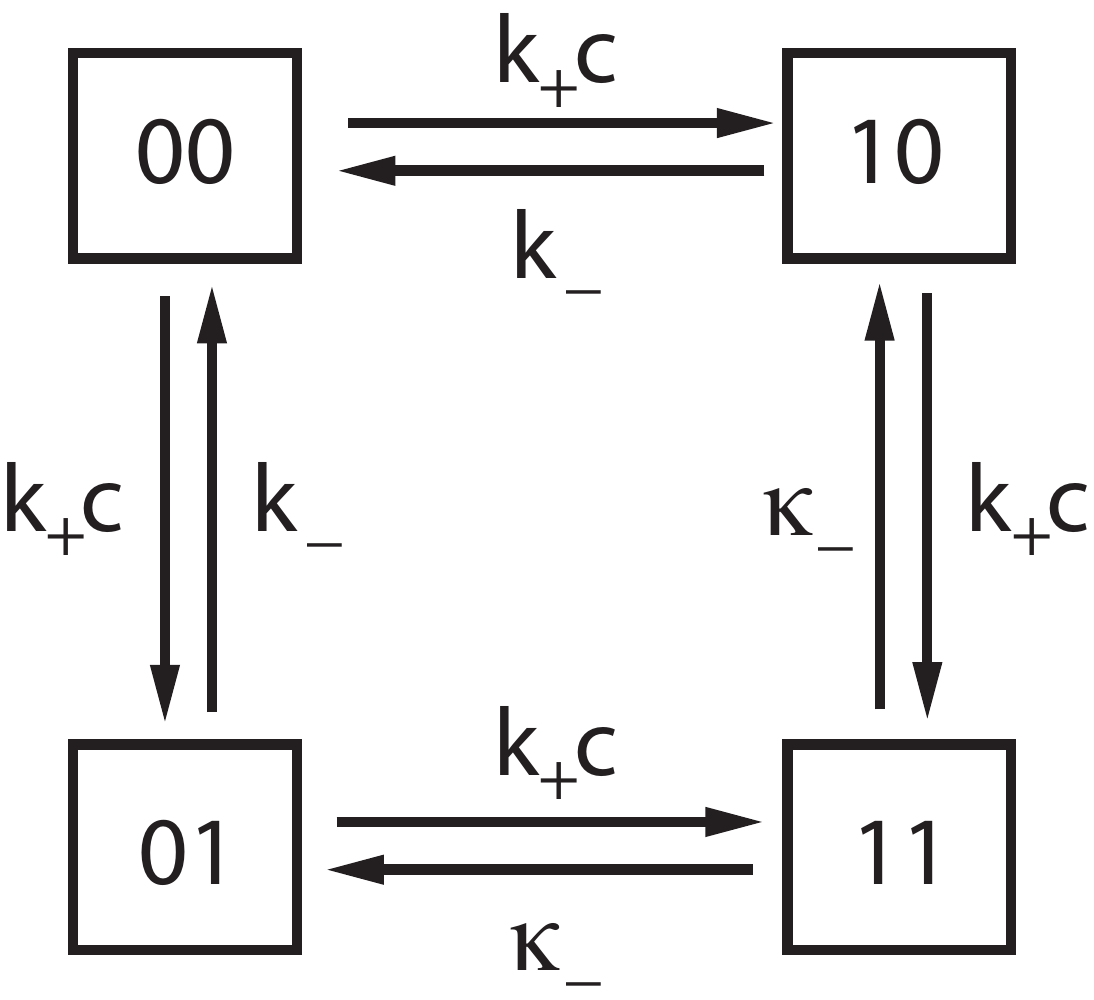}
\caption{The transitions in the model with 2 binding sites and 4 occupancy states, $(00,01,10,11)$. The binding of an additional molecule of TF happens at a rate $k_+c$, whereas unbinding rates are state dependent: a singly occupied promoter returns to the non-occupied state with a rate $k_-$, but the doubly occupied promoter loses a molecule of TF with the  rate $\kappa_-$. This difference is due to cooperativity, where the binding of one molecule stabilizes the binding of the other, and this makes the unbinding rates state dependent. The ``active'' state is $(11)$ in the lower right corner. We leave it as an exercise for the reader to write down the dynamical equations $dn_{00}/dt=\dots$, $dn_{01}/dt=\dots$ etc, observe that $n_{00}+n_{01}+n_{10}+n_{11}=1$, and compute the steady state activation if cooperativity is strong, $\bar{n}_{11}$. As in the case of a single binding site, this expression can be connected to the thermodynamic result of Eq (\ref{p11}). }
\label{f-2state}
\end{figure}

The probability of being active is then
\begin{equation}
P(11) = \frac{e^{-2E-\epsilon+2\mu}}{e^{-2E-\epsilon+2\mu} + 2e^{-E+\mu}+ 1}, \label{p11}
\end{equation}
where we use the units where $\beta=1$, that is, we express the energies and chemical potential in thermal units of $k_BT$.
If the cooperativity is strong, i.e. the additional gain in energy $\epsilon$ is larger than the favorable energy of putting a molecule of TF out of the solution onto the binding site, $\epsilon \ll \mu-E$, we can drop the middle term of the denominator in  Eq (\ref{p11}) and simplify it into:
\begin{equation}
P(11) = \frac{c^2}{c^2 + K_d^2}, \label{hill2}
\end{equation}
with $K_d=\exp[\beta (E+\epsilon/2)]$, where again we have used the definition of chemical potential $\mu$. This problem with 2 binding sites and 4 states of occupancy also has a complementary dynamical picture, which is already quite complicated, see Fig.~\ref{f-2state}. We also note that the same behavior for occupancy given by Eq~(\ref{hill2}) can be derived directly from a master equation, assuming that the binding of dimers is necessary to activate the gene ($k_+c^2 P_0(g)$ instead of $k_+c P_0(g)$ in Eq~(\ref{mastereq})).

\begin{figure}
\includegraphics[width =  \linewidth]{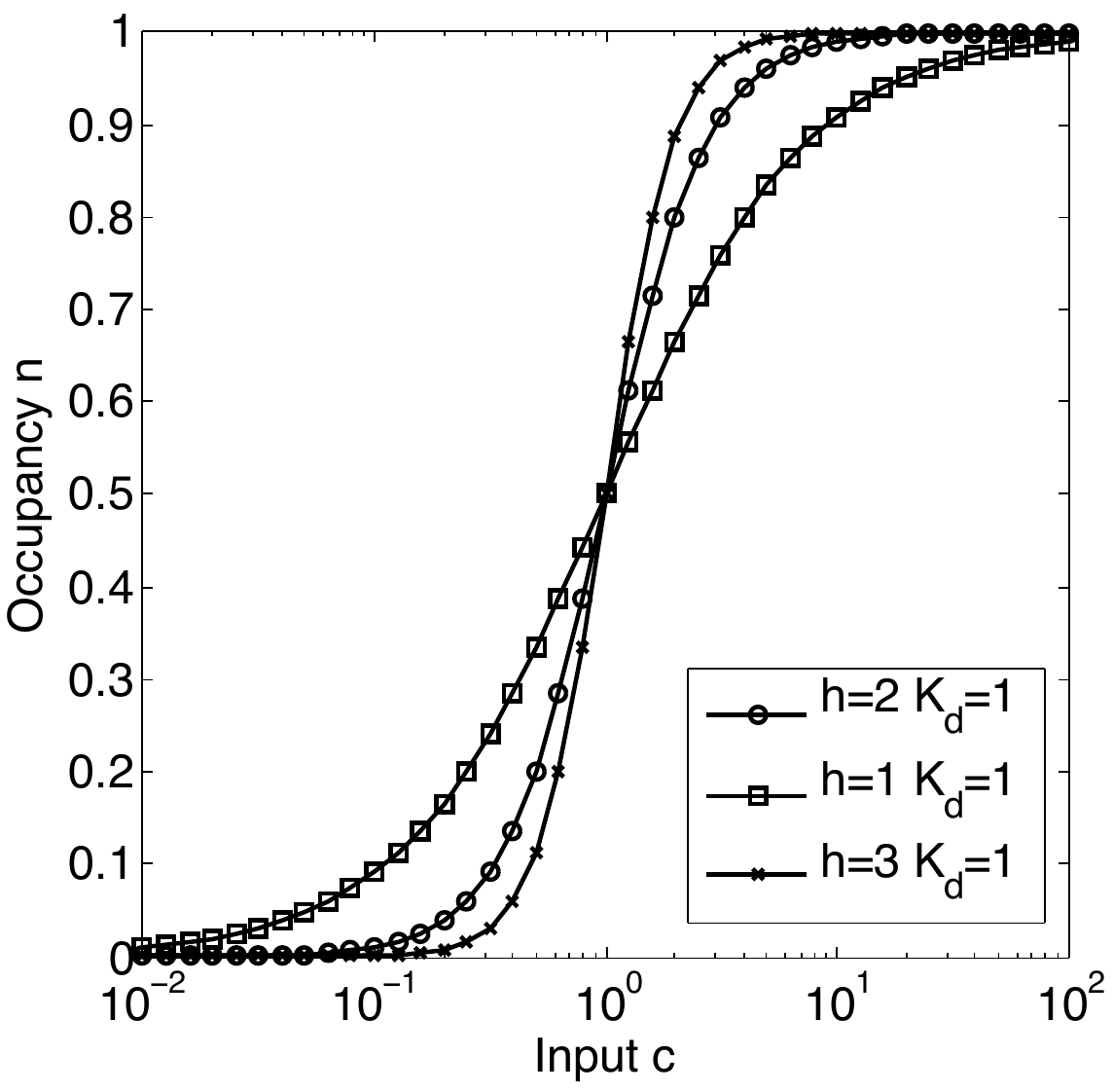}
\caption{Three Hill regulatory functions with different slopes (Hill coefficients $h$), as in the legend. All functions have $K_d=1$. Input TF concentration is customarily plotted on logarithmic horizontal axis, while the average promotor occupancy $\bar{n}$ is on the vertical axis. The output gene expression is in steady state $\bar{g}=(R\tau)\bar{n}(c)$, i.e. proportional to occupancy. The slope of $\bar{n}(c)$ on the log-log plot at half-induction ($c=K_d$) is related to the Hill coefficient, $d(\log \bar{n})/d(\log c)|_{K_d}=h/2$.}
\label{f-hill}
\end{figure}

Readers used to molecular biology models of gene regulation will recognize sigmoidal functions in Eqs (\ref{occaux},\ref{hill2}), also known as Hill functions, with a general form (see Fig.~\ref{f-hill}):
\begin{equation}
\bar{n}(c) = \frac{c^h}{c^h+K_d^h}, \label{hillg}
\end{equation}
where the dissociation constant $K_d$ is interpreted as the concentration at which the promoter is half induced, and $h$ is known as the cooperativity or Hill coefficient, usually interpreted as the ``number of binding sites''\footnote{In case where there is feedback regulation of the gene, for example through self-activation where a gene $g$ can activate its own transcription in addition to being activated by the input $c$, the interpretation of $h$ as the number of binding sites is incorrect.}.  Here we have shown how such phenomenological curves arise from simple statistical mechanics models of gene regulation with cooperative interactions. For repressors, one can show that $\bar{n}(c) = K_d^h/(c^h+K_d^h)$.

Before proceeding, let us inspect more closely the relation between the dynamical rates and the binding energy for a single site: $k_-/k_+=\exp(\beta E)$. As we have shown in Eq~(\ref{db}), this equality is required by detailed balance if thermodynamic and kinetic pictures are to match. Molecularly, the energy of binding $E$ in the case of transcription factor -- DNA interaction depends on the DNA sequence. So if we were to vary the sequence and binding energy $E$ would change, which of the two rates, $k_-$ or $k_+$ would vary as a result? In general one cannot answer this question without knowing in detail the sequence of molecular transitions that happen at the binding site. However, there is a useful limit, called the \emph{diffusion-limited on-rate}, that is often applicable. In this regime, the limit to how quickly a TF molecule can bind is given by the speed at which it can diffuse to the binding site. It has been shown that if a TF diffuses with diffusion constant $D$ and is trying to bind a site with linear dimension $a$, the fastest on-rate is $k_+\approx 4\pi D a$, for spherical TF and binding site\footnote{If assumptions about geometry are relaxed, the prefactor $4\pi$ will change.} \cite{bergreview}. In the diffusion-limited approach, if the binding site is empty, as soon as a TF diffuses into a region of size $a$ around the binding site, it will immediately bind. Then, all dependence on binding energy $E$ will be absorbed into the off-rate $k_-$. Intuitively we can understand this by imagining that once the TF is bound in an energetically favorable configuration, it has to wait for a random thermal kick of typical size $k_BT$ to unbind, and the probability of that kick being able to overcome the binding energy barrier $E$ is $\sim \exp(E/k_B T)$. We will return to this limit in Section \ref{Sourcenoise}.%
% cite Bintu et al
%
\subsection{Regulation by several transcription factors}
In the previous chapter we have shown how thermodynamic and kinetic models are connected for simple cases of gene regulation where a single transcription factor binds cooperatively to different numbers of binding sites. In many cases, however, several transcription factors together regulate a single gene. How can such situations be addressed from a theoretical perspective? We will describe two molecular frameworks for describing the joint regulation by two TFs. Both approaches can be easily generalized to more types of TFs. 

In the previous section we have motivated and derived Hill-type regulation functions. If we are considering a gene $g$ regulated by two TFs, we need to be precise how these proteins act together, that is, we need to specify the ``regulatory logic'' of their interaction. For example, if gene $g$ is activated by TF $A$, present at concentration $c_A$, and repressed by TF $B$, present at concentration $c_B$, one could postulate (without deriving) that the occupancy of the promoter is
\begin{equation}
\bar{n}(c_A,c_B)=\frac{c_A^{h_A}}{c_A^{h_A}+K_A^{h_A}}\cdot \frac{K_B^{h_B}}{c_B^{h_B}+K_B^{h_B}}. \label{andreg}
\end{equation}
This expression assumes that molecules of $A$ bind independently (of $B$) to $h_A$ sites with dissociation constant $K_A$, and molecules of $B$ bind to $h_B$ sites with dissociation constant $K_B$;  importantly, we also assume that the joint regulation is \emph{and}-like, meaning that gene $g$ will only be activated when both $A$ is bound and $B$ is \emph{not} bound [that's why there is a product in Eq (\ref{andreg})]. Conversely, in an alternative model the action of TF $A$ and TF $B$ could be additive:
\begin{equation}
\bar{n}(c_A,c_B)=\zeta_1\frac{c_A^{h_A}}{c_A^{h_A}+K_A^{h_A}}+ (1-\zeta_1)\frac{K_B^{h_B}}{c_B^{h_B}+K_B^{h_B}}. \label{indreg}
\end{equation}
$\zeta_1$ is a number between [0,1], which balances the effect of both types of TFs on the expression of gene $g$. Another model might assume a combination of cooperative regulation given by Eq~(\ref{andreg}) and an additive model given by Eq~(\ref{indreg}). More complex schemes like this one can clearly be derived, and while they will not necessarily correspond to any possible thermodynamic system, they might be useful \emph{phenomenological} models that can be fitted to the data. 

We can also pick a real thermodynamic model that is flexible enough to encompass many possible combinatorial strategies of gene regulation, while still having a small enough number of parameters to connect to available data. As in the previous case, this model might not correspond on a molecular level to the events on the promoter, and would thus also qualify as a phenomenological model. It would, however, have the advantage of being more easily interpretable and understandable within the context of statistical physics. One such model is the so-called Monod-Wyman-Changeaux (MWC) model.

The  MWC can easily be extended to include combinatorial regulation. The model has been motivated by the work on allosteric transitions and was used to explain hemoglobin function \cite{mwc}. When applied to the case of gene regulation, the central idea is that as a whole, the promoter can be in two states, ``on'' (1) and ``off'' (0). Remember that in our previous examples we had to declare one of the combinatorial states as the ``active'' state; here, this distinction is built into the model by assumption. See Ref~\cite{mirnymwc} for recent work that uses MWC to include the effect of nucleosomes on gene expression.

\begin{figure}
\includegraphics[width =  \linewidth]{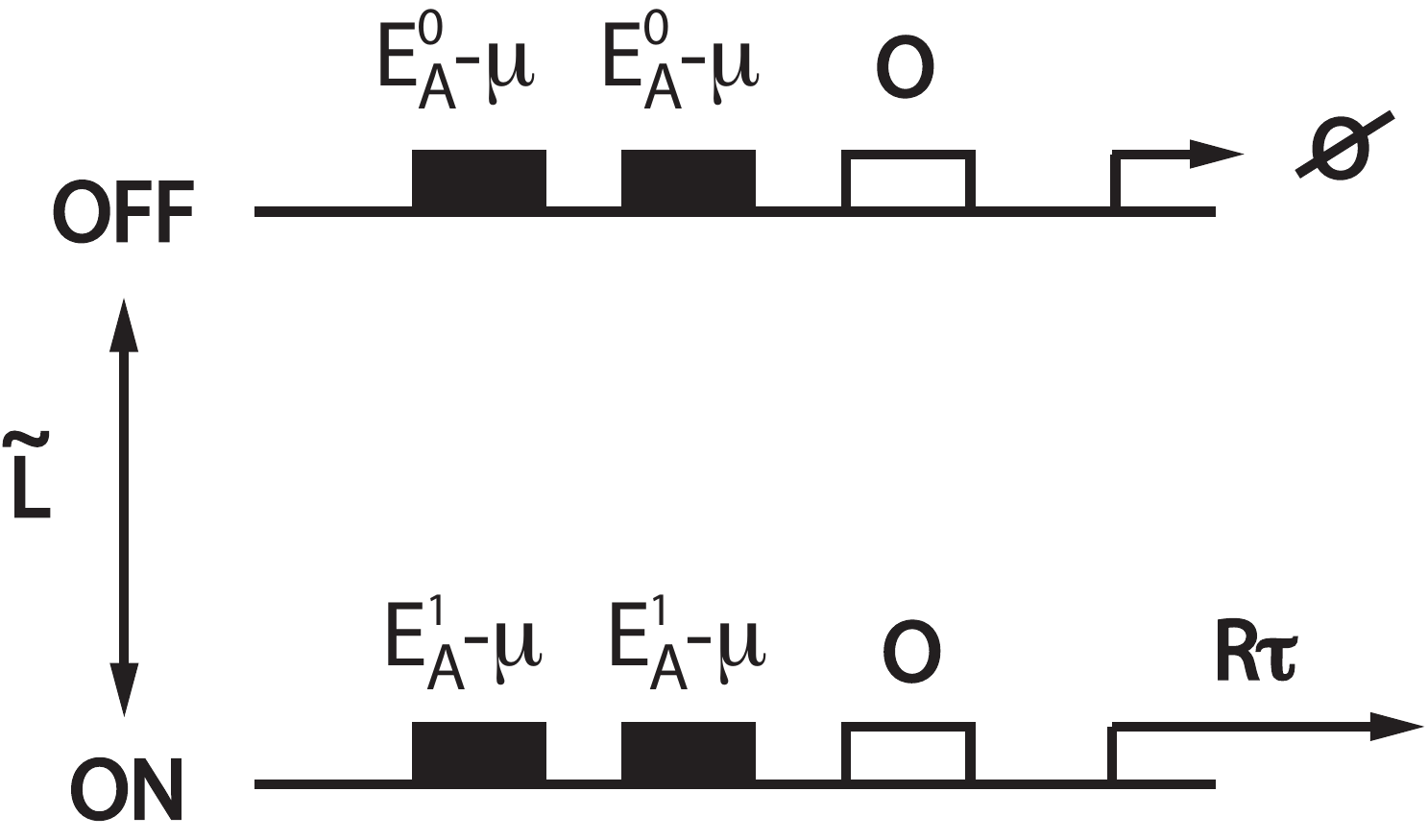}
\caption{A schematic diagram of MWC model. Two possible states of the promoter, ``on'' and ``off'', are separated by an energy barrier of $\tilde{L}$. There are 3 binding sites for the transcription factor in this example, to which TFs bind independently; their binding energy, however, depends on the state of the promoter. Here, 2 of the 3 sites are occupied, and are contributing $E_A^{0,1}-\mu$ each to the total free energy. If the promoter is ``off,'' there is no transcription, if it is ``on,'' transcription proceeds at rate $R$ and gives rise to $R\tau$ molecules of output in steady state at full induction.}
\label{f-mwc}
\end{figure}

The regulatory region has $n_A$ binding sites for transcription factor $A$. These sites can be bound in both the active and inactive state, and molecules of $A$ always bind independently, see Fig.~\ref{f-mwc}. However, the binding energy for each molecule of $A$ to its binding site is state-dependent, i.e. $E_A^0$ when the whole promoter is ``off'' vs $E_A^1$ when it is ``on.'' Let's work out the thermodynamics of this system. For each of the two states, we can write down the free energies of $k$ molecules of type $A$ bound:
\begin{eqnarray}
F_0&=&k(E_A^0 - \mu) + \tilde{L}	\\
F_1&=&k(E_A^1 - \mu), 	
\end{eqnarray}
where $\mu=\log c$ (we are writing everything in units of $k_BT$ and dimensionless concentration again), and $\tilde{L}$ measures how favoured the ``off'' state is against ``on'' state even with no TF molecules bound. The partition function is then
\begin{equation}
Z=\sum_{k=0}^n {n\choose k}e^{-k(E_A^0-\mu)+\tilde{L}} +\sum_{k=0}^n {n\choose k}e^{-k(E_A^1-\mu)}.
\end{equation}
Recognizing that the sums are simply binomial expansions\footnote{Note that $\sum_{k=0}^n{n \choose k}r^k=(1+r)^n$.}, we get for the probability of the ``on'' state (proportional to the expression of the gene):
\begin{eqnarray}
P(\mathrm{on}) &=& \frac{(1+e^{-E_A^1+\mu})^n}{(1+e^{-E_A^1+\mu})^n + (1+e^{-E_A^0+\mu})^ne^{\tilde{L}} } \label{mwce}\\
&=&\frac{(1+c/K_A^1)^n}{(1+c/K_A^1)^n+L(1+c/K_A^0)^n}. \label{mwcclass}
\end{eqnarray}
Equation (\ref{mwcclass}) is written in the standard form, with the identifications $K_A^1=\exp(\beta E_A^1)$, $K_A^0=\exp(\beta E_A^0)$ and $L=\exp(\tilde{L})$. 

The regulatory impact of transcription factor $A$ onto the regulated gene is described by quantities $(n,K_A^0,K_A^1)$ in the MWC model. There is one additional parameter $L$, the offset (or ``leak'') favoring the ``off'' state. Note that the parameters $K_A^{0,1}$ of the MWC model are not directly comparable to Hill model parameter $K_d$; however, we can make the identification in the regime where $c/K_A^0\ll 1$ and $c/K_A^1\gg 1$. Then the term $(1+c/K_A^0)^n$ in Eq~(\ref{mwcclass}) can be approximated with 1, and $(1+c/K_A^1)^n\approx (c/K_A^1)^n$. Equation~(\ref{mwcclass}) then reduces to
\begin{equation}
P(\mathrm{on})=\frac{c^n}{c^n + L(K_A^1)^n},
\end{equation}
and we can identify the parameter $n$ in the MWC model with the Hill coefficient $h$, and the dissociation constant of the Hill model, $K_d$, with $K_d=L^{1/n}K_A^1$.

In general, for a single gene, the MWC model is not much different from Hill functions, producing sigmoidal curves that don't necessarily cover the whole range from 0 to 1 in induction as the input changes over a wide range. However, in the limit where $c/K_A^0\ll 1$, we can easily generalize MWC to regulation by several transcription factors. To see how, rewrite Eq~(\ref{mwce}) as
\begin{equation}
P(\mathrm{on}) = \frac{1}{1+e^{F(c)}}, \label{fenergy}
\end{equation}
where $F(c) = -n \log(1+c/K_A^1) + \tilde{L}$. In this picture, the binding and unbinding of transcription factors simply shifts the free energy of ``on'' vs ``off'' state. We can easily see that if $K$ transcription factors $\mu = A, B, \dots$  with concentrations $c_\mu$ regulate the expression of a gene, we can retain Eq~(\ref{fenergy}), but write
\begin{equation}
F(\{c_\mu\})=-\sum_{\mu} n_{\mu}\log\left(1+\frac{c_\mu}{K_\mu}\right) + \tilde{L};
\end{equation}
it is easy to check that positive $n_\mu$ represent activating influences, while flipping the sign of $n_\mu$ makes that gene $\mu$ repress the expression of $g$ \cite{wtb_pre10}.

To summarize, different functional models of gene regulation presented in this section result in the steady-state output concentration of the gene, $\bar{g}(\{c_\mu\})\propto P(\mathrm{on})$, being a function of the concentration its TFs, $\{c_\mu\}$. We can think of these functions as nonlinear input/output relations, $\bar{g}=\bar{g}(\{c_\mu\})$ that can be computed theoretically and, in many cases, mapped out experimentally \cite{settyalon, kuhlmanhwa}.

\subsection{Sources of noise in gene expression}\label{Sourcenoise}

So far we have described several functional models for transcriptional regulation, and have shown how steady-state input/output relations, $\bar{g}=\bar{g}(\{c\})$, can be derived from kinetic and thermodynamic considerations. However, as we mentioned in the Introduction, gene expression is a stochastic process. What does this mean? In short, it means that the \emph{mean} input/output relations are not a full description of the system. Given an input $c$, the output $g$ will on average have the value $\bar{g}(c)$, but will dynamically fluctuate around this average. 

Alluding already to the terminology we are going to introduce more properly when discussing information transmission, we can view a genetic regulatory element as a ``channel'' that takes inputs $c$ and maps them into outputs $g$. When we say that there is noise in this mapping, we mean that for a single value of the input $c$, the output is not uniquely determined.  Instead, there exists a distribution over $g$, $P(g|c)$, that tells us how likely we are to receive a particular $g$ at the output if the symbol $c$ was transmitted. This distribution, $P(g|c)$, can be referred to as the conditional distribution of responses given the inputs. Once we know this distribution we can calculate (for continuous variables, such as concentrations), the mean response,  $\bar{g}(c)$, and the spread around the mean, characterized by the variance $\sigma_g^2(c)$:
\begin{eqnarray}
\bar{g}(c) &=& \int dg\;g P(g|c)	\label{condmean},\\
\sigma_g^2(c)&=&\int dg\;(g-\bar{g})^2 P(g|c). \label{condvar}
\end{eqnarray}
These two functions are known as conditional mean and conditional variance, and they can easily be extracted from the distribution $P(g|c)$, if it is known. A noise-free deterministic limit is recovered as $\sigma_g^2(c)\rightarrow 0$, in which case $P(g|c)$ tends to a Dirac-delta distribution, $P(g|c)=\delta(g-\bar{g}(c))$.

Unfortunately, the full conditional distribution of responses given the inputs, $P(g|c)$, is usually only available in theoretical calculations or simulations, since in reality we rarely have enough data to sample it. In the case of gene regulation, sampling would involve changing the input concentration of TF, $c$, and for each input concentration, measuring the full distribution of expression levels $g$. More often than not we only have enough samples to measure a few moments of the conditional output distribution, perhaps the conditional mean and conditional variance. Given these measurements and $P(g|c)$ that is experimentally inaccessible directly by sampling, we can try making the approximation
\begin{equation}
P(g|c)\approx \mathcal{G}(g; \bar{g}(c), \sigma_g^2(c)),
\end{equation}
that is, we \emph{assume} that $P(g|c)$ is a Gaussian, with some input-dependent mean and variance.
In the presented setting, the mean input/output response and the noise in the response cleanly separate: one is given by the conditional mean, and the other by conditional variance. The noise can be thought of as \emph{the fluctuations in the output variable while the input is held fixed}. Recall that we are discussing all information processing systems in equilibrium, that is, when the dynamics in $g$ has reached steady state (and all variation in $g$ at given $c$ is due to noise). Having built these intuitions, let us see how noise can be derived in a simple model of gene regulation. 

\subsection{Derivation of noise for simple gene regulation}\label{Dernoise1}
To start, we first return to the simple gene regulation scenario of Fig.~\ref{f-sscheme}. We will sketch how the noise can be derived in this model using the Langevin approximation, and give a back-of-the envelope estimate for the terms that we do not compute here. The reader is invited to view the full derivation in Ref~\cite{tkacik+al_08a}.%cite

We start with the dynamical equations:
\begin{eqnarray}
\frac{dn}{dt}&=& k_+c(1-n) -k_-n + \xi_n	\label{noise1}\\
\frac{dg}{dt}&=& Rn - \frac{1}{\tau}g + \xi_g, \label{noise2}
\end{eqnarray}
where again we take the binding site occupancy $n$ to be between 0 and 1, and the expression level of the output gene is $g$; $g$ is produced with rate $R$ when the binding site is occupied, and the proteins have a lifetime of $\tau$. We have already shown that the equilibrium solution of this system is $\bar{n} = k_+c/(k_+c + k_-)$ and $\bar{g} = (R\tau) \bar{n}$. Here we are interested in the fluctuations, $\sigma_g(c)$, around the steady state, that arise purely due to intrinsic noise sources: {\bf (i)} the fact that the binding site only has two binary states that switch on some characteristic timescale, {\bf (ii)} the fact that we make a finite number of discrete proteins at the output, and {\bf (iii)} the fact that the input concentration $c$ might itself fluctuate at the binding site location.

\begin{figure}
\includegraphics[width =  \linewidth]{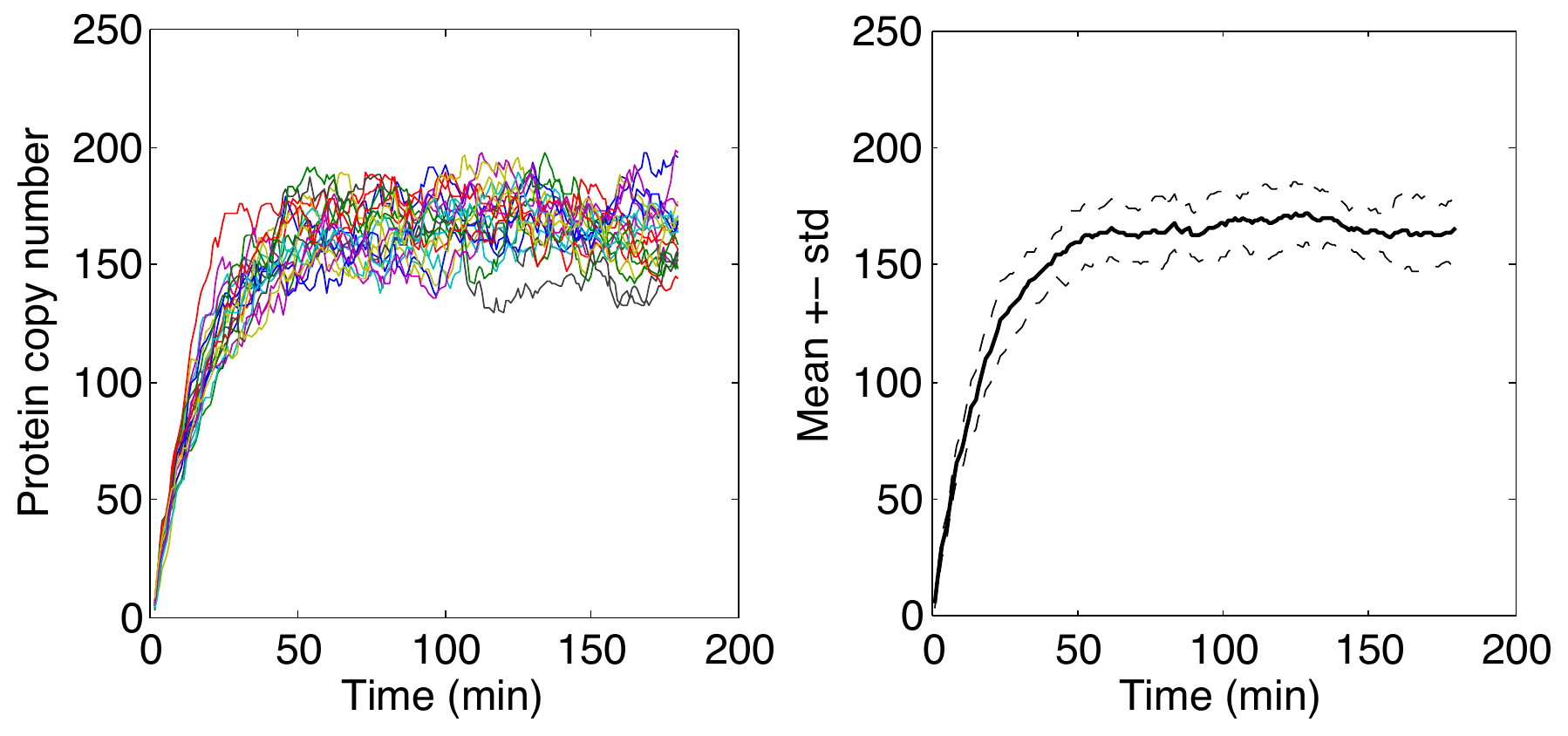}
\caption{A fully stochastic simulation of a simple model of gene expression using reactions specified in Eq~(\ref{cgillespie}). The simulation starts with $g(t=0)=0$ proteins; the steady state is reached after about 70 minutes. On the left, the trajectories of 20 simulation runs. On the right, the mean trajectory plotted in a solid line; mean $\pm$ 1-std plotted in dashed lines. The envelope measures the steady state level of noise due to (i) random promoter switching and (ii) the shot noise in producing the output molecules. }
\label{f-gillespie}
\end{figure}

One approach would be to simulate the system of Eqs~(\ref{noise1},\ref{noise2}) exactly using the Gillespie SSA algorithm \cite{Gillespie}. For a given and fixed level of input $c$, the results of 20 such simulation runs are shown in Fig.~\ref{f-gillespie}.

To compute this noise analytically instead of using the simulation, we have introduced random Langevin forces $\xi_n$, $\xi_g$. Consider the second equation, Eq~(\ref{noise2}). A single protein is produced anew, or is degraded, as an elementary step (since you don't make half a protein). In equilibrium, the production term $R\bar{n}$ balances the degradation term, $\bar{g}/\tau$. Now consider some time $T$ in which $RT\bar{n}=\bar{g}T/\tau \approx 1$, i.e. one molecule is produced or destroyed on average and with equal probability. While the expected change in the total number in equilibrium in time $T$ is zero, the variance is not: the variance is equal to $\frac{1}{2}\times$ (production of 1 molecule)$^2$ + $\frac{1}{2}\times$ (degradation of 1 molecule)$^2$ = 1. In general, the variance will be $T(R\bar{n} + \bar{g}/\tau)$ if we measure for time $T$. If you are familiar with random walks in 1D, this sounds very familiar: the mean displacement is 0 (because ``leftwards steps'' = steps that decrease protein copy number, and ``rightwards steps'' = steps that increase protein copy number, are equally likely), but the variance in displacement from the origin grows with time $T$. 

Statistical physics tells us that in order to reproduce this variance in a dynamical system, we have to insert Langevin forces with the following prescription:
\begin{eqnarray}
\langle \xi_g(t)\rangle &=& 0	\nonumber\\
\langle \xi_g(t) \xi_g(t')\rangle &=& (R\bar{n} + \bar{g}/\tau) \delta(t-t'). \label{lngforce}
\end{eqnarray}
The mean random force is zero, it is uncorrelated in time, and it has an amplitude such that the random kicks have variance equal to the leftward and rightward step size; this will recover our intuition about 1D random walks. We note that the Gaussian assumption holds for large copy numbers and that at very short timescales the assumption of temporally uncorrelated noise can break down. Similarly, $\langle\xi_n(t)\xi_n(t')\rangle=(k_+c(1-\bar{n})+k_-\bar{n})\delta(t-t')$. 

To proceed, we first linearize Eqs~(\ref{noise1},\ref{noise2}) around the equilibrium, by writing $n(t)=\bar{n}+\delta n(t)$, $g(t)=\bar{g}+\delta g(t)$. Then we introduce Fourier transforms:
\begin{eqnarray}
\delta n(t)& =& \int \frac{d\omega}{2\pi} \delta \tilde{n}(\omega) e^{-i\omega t}\\
\delta g(t)&=& \int \frac{d\omega}{2\pi}\delta \tilde{g}(\omega) e^{-i\omega t}.
\end{eqnarray}
Fourier transforms of $\xi_n$ and $\xi_g$ are simply $\tilde{\xi}_n=2k_-\bar{n}$ and $\tilde{\xi}_g=2R\bar{n}$, respectively (because the Fourier transform of a delta-function is 1, and we have also used the fact that in equilibrium, the two terms that contribute to the magnitude of each Langevin force are equal). 

With this in mind, the system of equations in the Fourier space (denoted by tildes) now reads:
\begin{eqnarray}
-i\omega \delta \tilde{n}&=& -\frac{1}{\tau_c}\delta \tilde{n} + \tilde{\xi}_n \label{deriv1}\\
-i\omega\delta\tilde{g}&=&R\delta\tilde{n}- \frac{1}{\tau}\delta\tilde{g} + \tilde{\xi}_g, \label{deriv1a}
\end{eqnarray}
where $\tau_c^{-1}=(k_+c+k_-)$. 

We ultimately want to compute $\sigma_g^2(c)$. The total variance is composed from fluctuations at each frequency $\omega$, integrated over frequencies\footnote{Because the noise process is stationary (time-translation invariant), the noise covariance $\langle \delta g(t)\delta g(t')\rangle=C_g(|t-t'|)$ will depend on the difference in time only, and going into the Fourier basis will diagonalize the covariance matrix. The total noise variance is the integral over these independent Fourier components, and that is equal by Parseval's theorem to the total noise variance obtained by doing the corresponding integral in the time domain. }:
\begin{equation}
\sigma_g^2=\int \frac{d\omega}{2\pi}\langle \delta\tilde{g}(\omega) \delta\tilde{g}^*(\omega)\rangle = \int\frac{d\omega}{2\pi}S_g(\omega), \label{ngg}
\end{equation}
where $S_g(\omega)$ is called the \emph{noise power spectral density} of $g$, and the asterisk denotes complex conjugate.
We see that we need to solve for $\delta\tilde{g}$ first from Eqs.~(\ref{deriv1},\ref{deriv1a}):
\begin{eqnarray}
\delta\tilde{g} = \frac{R\tilde{\xi}_n}{(-i\omega + \tau_c^{-1})(-i\omega+\tau^{-1})} + \frac{\tilde{\xi}_g}{-i\omega+\tau^{-1}}.
\end{eqnarray}
Next, we compute $\langle \delta\tilde{g}(\omega) \delta\tilde{g}^*(\omega)\rangle$. Recalling the definitions of $\langle \tilde{\xi}\tilde{\xi}^*\rangle$ [Eq~(\ref{lngforce})], we find that 
\begin{eqnarray}
S_g(\omega) &=& \frac{R^2(2k_-\bar{n})}{(\omega^2+\tau_c^{-2})(\omega^2+\tau^{-2})} + \frac{2R\bar{n}}{\omega^2+\tau^{-2}}.
\end{eqnarray}
The binding and unbinding of the promoter is usually much faster than the protein decay time, $\tau_c\ll \tau$. Using this and the fact that $\int_{-\infty}^{\infty}dx (x^2 + 1)^{-1}=\pi$, we finally find
\begin{equation}
\sigma_g^2(c) = \bar{g}(c) + \frac{(R\tau)^2}{k_-\tau}\bar{n}(1-\bar{n})^2. \label{rawnoise}
\end{equation}
If we normalize the expression level $g$ such that it ranges between 0 (no induction) to 1 (full induction) by defining $\hat{g}=\bar{g}/(R\tau)$, then the noise in $\hat{g}$ is 
\begin{equation}
\sigma_{\hat{g}}^2(c) = \frac{1}{R\tau}\hat{g} + \frac{1}{k_-\tau}\hat{g}(1-\hat{g})^2. \label{fnoise1}
\end{equation}

Our result is lacking at least one important contribution to the total  noise. The formal derivation of this missing term is involved \cite{simapaper, tkacik+al_08a}, so we will estimate it here up to a prefactor. In our derivation we have not taken into account that the molecules of transcription factor are brought to the binding site by diffusion. The diffusive arrival of molecules into a small volume around the binding site is a random process as well: it will induce some noise in occupancy of the binding site, and thus in the expression level $\hat{g}$. This is the contribution we are going to estimate.

Suppose that the binding site is fully contained in a physical box of side $a$. When the average TF concentration in the nucleus is fixed at $\bar{c}$, the average number of molecules in the box is $\bar{N}=a^3\bar{c}$. This, however, is only the mean number; if we were to actually sample many times the number of molecules  in the box, we would find that our counts are distributed in a Poisson fashion, with a variance equal to the mean: $\sigma_N^2=\bar{N}$. This is again just the familiar shot noise, now appearing at the input side.

How can one reduce the fluctuations $\sigma_N^2$? As always, one can make more independent measurements, and average the noise away. With $M$ independent measurements, the effective noise should decrease, $\sigma_{N,\mathrm{eff}}^2=\sigma_N^2/M$. Suppose the binding site measures for a time $\tau$ (the protein lifetime, the longest time in the system). How many independent measurements  were made in the best possible case? It takes $t_0=a^2/D$ time for the molecules to diffuse out of the box of size $a$ and be replaced with new molecules; if we take snapshots and count the molecules at intervals faster than $t_0$, we are not making independent measurements. Therefore $M=\tau/t_0=\tau D/a^2$. Plugging this into the expression for effective noise, we find $\sigma_{N,\mathrm{eff}}^2=a^3\bar{c}\times a^2/(D\tau)$. Since $\bar{N}=a^3\bar{c}$, it follows that $\sigma_N^2=a^6\sigma_c^2$, and finally:
\begin{equation}
\sigma_{c,\mathrm{eff}}^2=\frac{\bar{c}}{Da\tau}. \label{cnoise}
\end{equation}
Equation (\ref{cnoise}) is a fundamental result: any detector  of linear size $a$ measuring concentration $c$, to which ligands are transported by diffusion with coefficient $D$, and making measurements for time $\tau$, will suffer from the error in measurement in concentration, given by $\sigma_c$. This contribution to the noise is called \emph{diffusive noise}, and it is a special form of input noise.

\begin{figure}
\includegraphics[width =  \linewidth]{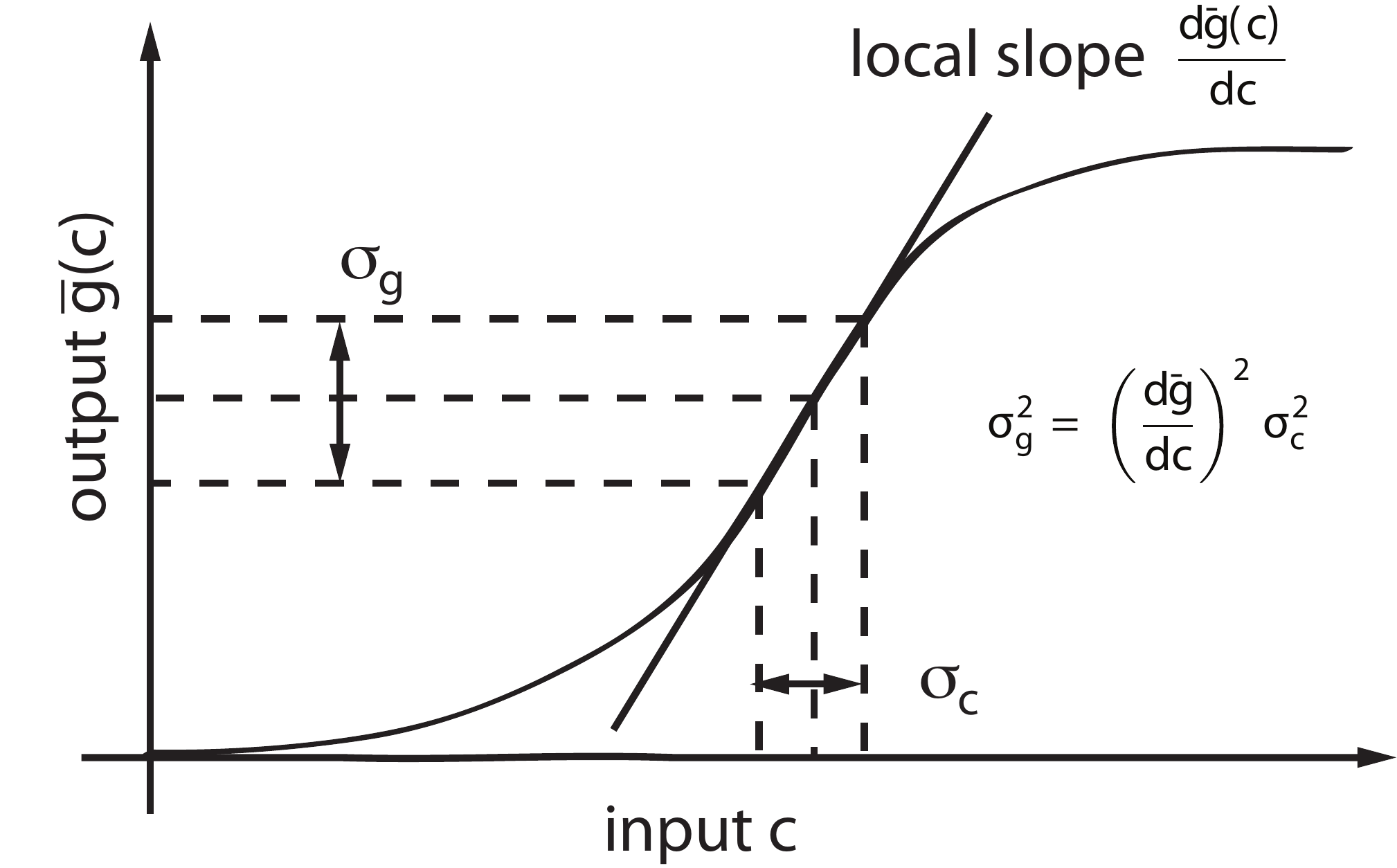}
\caption{Propagating the noise in the input $\sigma_c$, through the mean input/output relation, $\bar{g}(c)$, into the effective noise in the output, $\sigma_g$. The variances are related by the square of the local slope of the input/output curve, $d\bar{g}/dc$.}
\label{f-noiseprop}
\end{figure}

To assess how this input noise maps into the noise in the gene expression $g$, note that any (small) error at the input can be propagated to the output through the input/output relation, $\bar{g}(c)$ [see Fig~\ref{f-noiseprop}]:
\begin{equation}
\sigma_g^2=\left(\frac{d\bar{g}}{dc}\right)^2\sigma_c^2. 
\end{equation}
Adding the diffusive noise to previously computed terms in Eq~(\ref{fnoise1}), we find:
\begin{equation}
\sigma_{\hat{g}}^2(c) = \frac{1}{R\tau}\hat{g} + \frac{1}{k_-\tau}\hat{g}(1-\hat{g})^2 + \frac{\hat{g}^2(1-\hat{g})^2}{Da\bar{c}\tau}. \label{noisefinal}
\end{equation}

Let us stop here with the derivation, interpret the terms and summarize what we have learned so far. We tried to compute various contributions to the noise in the expression of gene $g$, in a simple regulatory element where the TF $c$ regulates $g$. In any real organism, such a small regulatory element will be embedded into the regulatory network, and $c$ will experience fluctuations on its own that will be transmitted into fluctuations in $g$, the so-called \emph{transmitted} noise, in addition to intrinsic noise calculated here \cite{pedraza}.

On top of intrinsic and transmitted noise sources, the output will also fluctuate due to the extrinsic noise because the cellular environment of the regulatory network is not stable. But even without these complications, we can identify at least three contributions intrinsic to the $c\rightarrow g$ regulatory process:

{\bf Output noise.} This is the first term in Eq~(\ref{noisefinal}), where the variance $\sigma_{\hat{g}}^2\propto \hat{g}$. Funamentally, this is a form of shot noise that arises because we produce a finite number of discrete output molecules. In the simple setting discussed here, the proportionality factor really is 1 [when $g$ is measured in counts, as in Eq~(\ref{rawnoise})], and this is a true Poisson noise where variance is equal to the mean. If we treated the system more realistically, with separate transcription and translation steps, the proportionality constant could be different from 1; a more careful derivation shows that then, $\sigma_{\hat{g}}^2=(1+b)/(R\tau)\hat{g}$, where $b$ is the burst size, or the number of proteins produced per single mRNA transcript, on average \cite{tkacik+al_08a}. This is easy to understand: the ``rare'' event is the transcription of a mRNA molecule, and that has true Poisson noise statistics, but for each single mRNA the system produces $b$ proteins, and the variance is thus multiplied by $b$.

{\bf Input promoter switching noise.} This is the second term in Eq~(\ref{noisefinal}). The source of this noise is binomial switching of the promoter, as it can only be in an induced ($n=1$) or empty ($n=0$) states. If we interpret $\bar{n}$ as the probability of being occupied, then the variance must be binomial $\bar{n}(1-\bar{n})$. Fluctuations between empty and full states of occupancy happen with the timescale $\tau_c$ [see Eq~(\ref{deriv1})], and the system averages for time $\tau$, so $\tau/\tau_c$ independent measurements are made, reducing the binomial variance to $\bar{n}(1-\bar{n})\tau_c/\tau$. Since $\tau_ck_-=(1-\bar{n})$ and $\bar{n}=\hat{g}$, we recover the switching term, $\hat{g}(1-\hat{g})^2/(k_-\tau)$. 

This term depends on the microscopic way the promoter is put together, hence the dependence on the kinetic parameter $k_-$. Regardless of these details, however, every promoter that has an ``on'' and ``off'' state will experience fluctuations similar in form to these derived here. In our example, $k_-$ is the rate of TF unbinding from the binding site and this is usually assumed to be very fast compared to the protein lifetime (in other words, the occupancy of the promoter is equilibrated on the timescale of protein production). In other scenarios that effectively induce gene switching, however, this assumption of fast equilibration might not be true. In particular, attention has lately been devoted to DNA packing and regulation via making the genes (in)accessible to transcription using chromatin modification. The packing / unpacking mechanisms are thought to occur with slow rates, and such switching term might be an important contribution to the total noise in gene expression \cite{rajplos}.

{\bf Input diffusion noise.} The last term in Eq~(\ref{noisefinal}), as discussed, captures the intuition that even with the fixed average concentration $\bar{c}$ in the nucleus (that is, even if $c$ did not undergo any fluctuation relating to its own production, degradation and regulation), there would still be \emph{local} fluctuations at its TF binding site location, causing noise in $g$. This contribution is important when $c$ is present at low concentrations. As an exercise, one can consider the approximate relevance of this term in case of prokaryotic transcriptional regulation, where $D\sim 1\e{\mu m}^2/\e{s}$, the size of the binding site $a\sim 3\e{nm}$, the relevant TF concentrations are in nanomolar range, and the integration times in minutes. It has been shown that this kind of noise also represents a physical limit in the sense that it is independent of the molecular machinery at the promoter, as long as the predominant TF transport mechanism is free diffusion.

\begin{figure}
\includegraphics[width =  \linewidth]{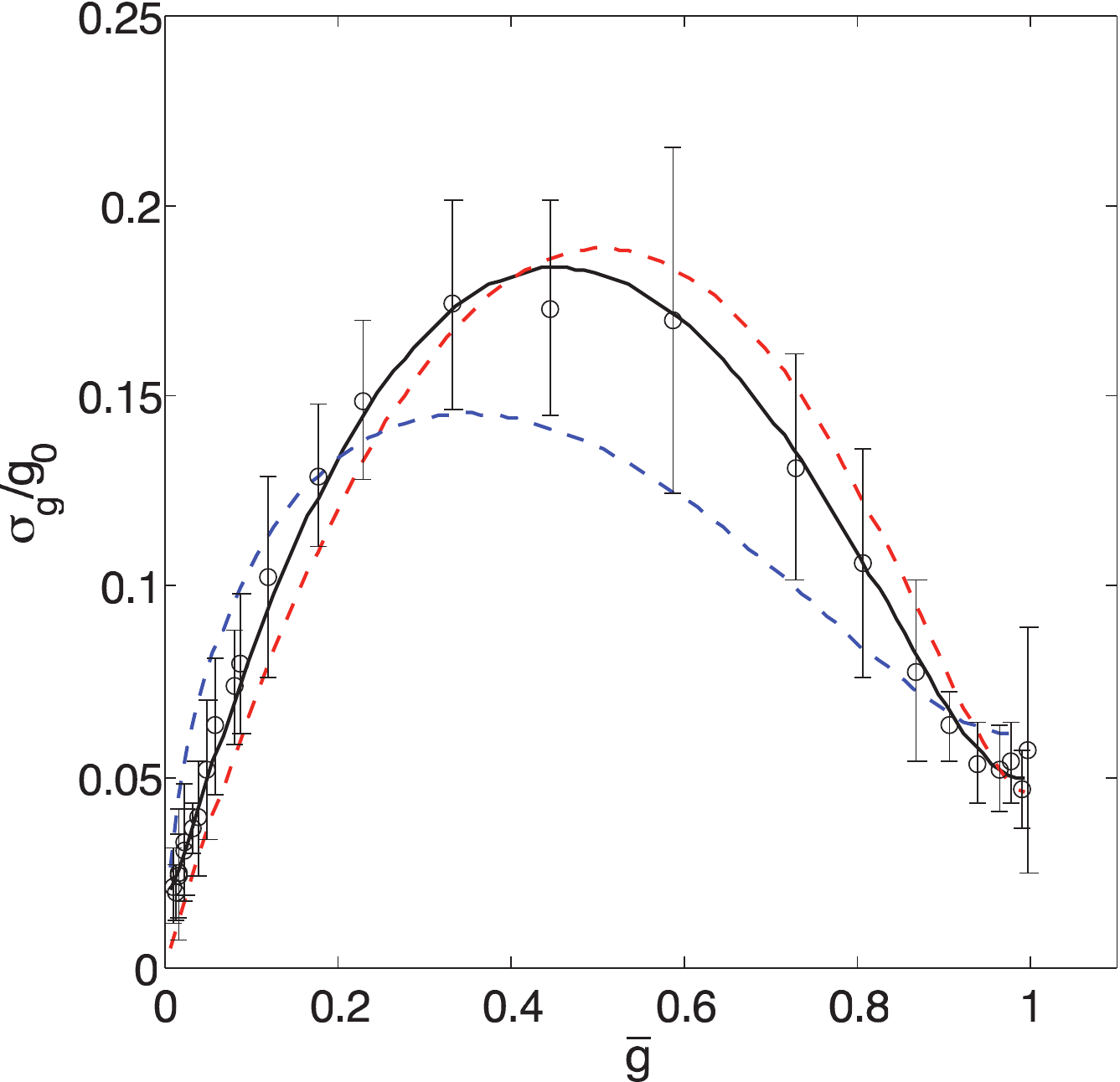}
\caption{The behavior of noise in hunchback expression, $\sigma_g$, as a function of the mean induction level of hunchback, $\bar{g}\in[0,1]$; reproduced from Ref~\cite{tkacik+al_08a} with data from Ref~\cite{Gregor}. Data points (black circles) show the measurement in 9 fly embryos at nuclear cycle 14; each point is an average across nuclei receiving the same input concentration of bicoid,  error bars are std across embryos. Solid lines show model fits: blue dashed line is a two-parameter fit with input switching and output noise contributions; red dashed line is a two parameter fit with input diffusion and output noise contributions, assuming step-like regulation of bcd/hb (infinite Hill coefficient); solid black line is a two-parameter fit with input diffusion and output noise contributions, with the mean input/output relation $\bar{g}(c)$ inferred from the data (Hill coefficient $\sim 5$). The black line is a very good fit to the data, indicating that the diffusion input noise contribution, responsible for the peak, is dominant, while the output noise contribution, responsible for the noise magnitude at full induction where $\bar{g}=1$, is smaller.  }
\label{noise5}
\end{figure}

What we presented here theoretically was a simple example, but how does it relate to experiment? Figure~\ref{noise5} shows that our simple model incorporating only the output  and the input diffusive noise contributions is an excellent description of data from early fly development. The two fitted parameters give the magnitudes of the two respective noise sources, and their values match the values estimated from known parameters and concentrations \cite{tkacik+al_08a}. We note that the prominent contribution of input noise seems to be a hallmark of noise in eukaryotic (but not prokaryotic) gene regulation.

Let us briefly summarize our observation about the noise: \\
{\bf (i)} Not only can we make models for \emph{mean} input/output relations $\bar{g}(\{c_\mu\})$, but we can compute the noise itself, as a function of the input, $\sigma_g(\{c_\mu\})$. Noise behavior is connected to the kinetic rates of molecular events, which are inaccessible in any  equilibrium measurement of mean input/output behavior. Therefore, if noise is experimentally accessible, it provides a powerful complementary source of information about transcriptional regulation. \\
{\bf (ii)} There are fundamental (physical) sources of noise which biology cannot avoid by any ``clever'' choice of regulatory apparatus; thus the precision of every regulatory process must be limited. These sources all fundamentally trace back to the finite, discrete and stochastic nature of molecular events. In theory, the corresponding noise terms thus have simple, universal forms, and we can hope to measure them in the experiment.  \\
{\bf (iii)} There are sources of noise in addition to the fundamental, intrinsic ones, including extrinsic, experimental, etc. The hallmark of a good experiment is the ability to separate these sources by clever experimental design and/or analysis; see e.g. Refs~\cite{elowitztwo,Gregor}.

\section{Introduction to information theory}
\subsection{Statistical dependency}\label{info_why}
Up to this point we have stressed the role of noise in biological networks and mentioned several times that noise limits the ability of the network to transmit information; in this section we will turn this intuition into a mathematical statement.

Recall that in our introduction to noise, we started with a probabilistic description of an information transmission system: given some input $c$, the system will map it into the output $g$ using a probabilistic mapping, $P(g|c)$. In case there were no noise, there would be no ambiguity, and $g=g(c)$ would be a one-to-one function.

Suppose that the inputs are drawn from some distribution $P(c)$ and fed into the system which responds with the appropriate $g$. Then, pairs of input/output symbols are distributed jointly according to
\begin{equation}
P(c,g)=P(g|c)P(c)	\label{joint}
\end{equation}
In what follows, we will be concerned with finding ways to measure how strongly the inputs ($c$) and the outputs ($g$) are dependent on each other. It will turn out that the general measure of interdependency will be tightly related to the concept of \emph{information}.

Let's suppose that our information transmission ``black box'' would be a hoax, and instead of encoding $c$ into $g$ in some fashion, the system would simply return a random value for $g$ no matter the input $c$. Then $c$ and $g$ would be \emph{statistically independent}, and $P(c,g)=P(c)P(g)$; such a box could not be used to transmit any information. As long as this is not true, however, there will be some statistical relation between $c$ and $g$, and we want to find a measure that would quantify ``how much'' one can  know, in principle, about the value of $c$ by receiving outputs $g$, given that there is some input/output relation $P(g|c)$ and some distribution of input symbols $P(c)$. 

\begin{figure}
\includegraphics[width =  \linewidth]{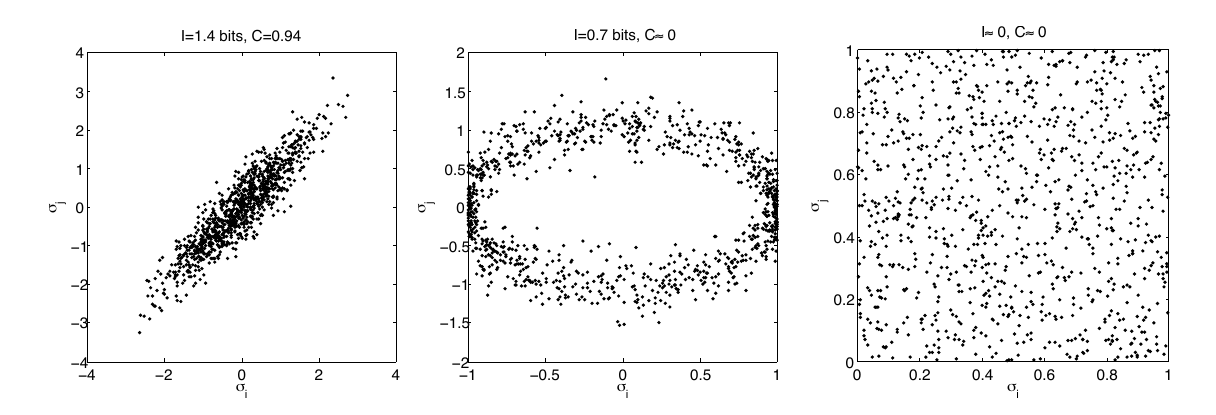}
\caption{Examples of two variables, drawn from three joint distributions. Shown are the scatterplots of example draws. On the left, the variables are linearly correlated, and the correlation is close to 1. In the middle, the variables are interdependent, but not in a linear sense. The correlation coefficient is 0, but measures of statistical dependence, such as \emph{mutual information}, give non-zero value. Note that we are looking for a general measure of interdependency: if we had a model that \emph{assumes} that $x$ and $y$ lie on a circle, we could fit that particular model or use a measure that makes the circular assumption. Instead, we would like to find a measure that detects the dependency without making any assumptions about the distribution from which the data has been drawn. On the right, the variables are statistically independent, and both linear correlation and mutual information give zero signal.   }
\label{scall}
\end{figure}

The first quantity that comes to mind as the interdependency measure between $c$ and $g$ is just the covariance:
\begin{eqnarray}
\mathrm{Cov}(c,g)=\int dc \int dg (c-\bar{c})(g-\bar{g}) P(c,g);
\end{eqnarray}
it is not hard, however, to construct cases in which the covariance is 0, yet $c$ and $g$ are statistically dependent. Covariance alone (or correlation coefficient) only tells us about whether $c$ and $g$ are \emph{linearly related}, but there are many possible nonlinear relationships that covariance does not detect; for example, see Fig~\ref{scall}.

Moreover, we would like our dependency measure to be very general (free of assumptions about the form of the probability distribution that generated the data) and definable for both continuous as well as discrete outputs\footnote{Covariance can be problematic when used on discrete quantities.}. We will claim, following Shannon \cite{shannon_48}, that there is a unique assumption-free measure of interdependency, called the \emph{mutual information} between $c$ and $g$. First, let us build some intuition.

\subsection{Entropy and mutual information}\label{info}
\begin{figure}
\includegraphics[width =  \linewidth]{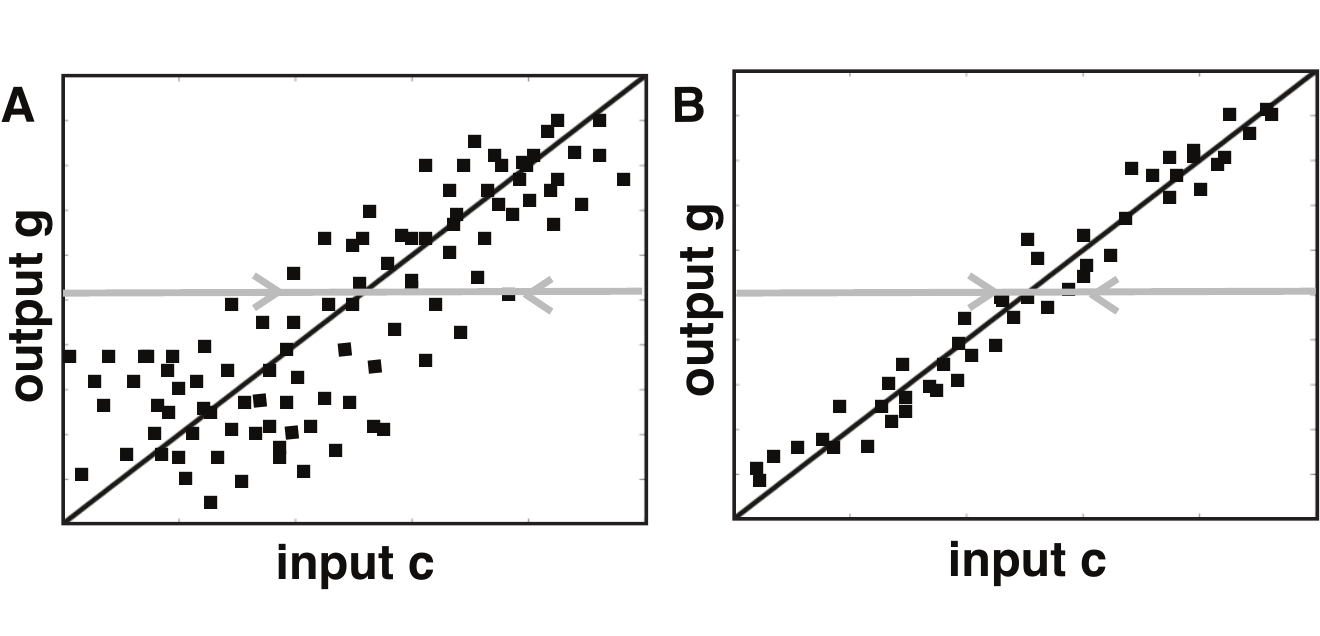}
\caption{A schematic depiction of two mock measurements (dots) of an output $g$ as a function of input $c$.  {\bf A)} A case where measuring the output does not greatly decrease our uncertainty about the input. This input/output relation has little information. {\bf B)} In this case the input/output relation is informative: measuring the output significantly reduces our uncertainty about the input. The grey line denotes a chosen value of the output, and the arrows mark the uncertainty in the input for that chosen value of the output.}
\label{intuitiveinfo}
\end{figure}

In a gene regulatory network, the concentrations of the input regulatory signal $c$ and the output effector protein $g$ are (nonlinearly) related through some noisy input/output relation. We want to consider how much information the input signal conveys about the level of the output. In general, we have an intuitive idea of information, which is schematized in Fig~\ref{intuitiveinfo}. In an experiment we could measure pairs of  $(c,g)$ values while the network performs its function, and scatterplot them as in Fig~\ref{intuitiveinfo}. The line represents a smooth (mean) input/output relation and guides our eyes. In the case of the mock measurements in Fig~\ref{intuitiveinfo}A, knowing the value of the output would tell us only a little about which value of the input generated it (or vice versa -- knowing the input constrains the value of the output quite poorly). However in the case of the input/output relation in Fig~\ref{intuitiveinfo}B, knowing the value of output would reduce our uncertainty about the input by a significant amount. Intuitively we would be led to say that in ``noisy'' case A there is a small amount of information between the input and the output, while in case B there is more.  From this example we see that information about $g$ obtained by knowing $c$ can be viewed as a ``reduction in uncertainty'' about $g$ due to the knowledge of $c$.  In order to formalize this notion we must first define uncertainty, which we do by means of the familiar concept of entropy.

Physicists often learn about entropy in the microcanonical ensemble, where it is simply defined as a measure of how many states are accessible in an isolated system at fixed energy, pressure and particle number. In this case all, say $M$, states that the system can find itself in, are equally likely, therefore the probability distribution $p_i$ over a set of states $i$,  such as the particle configurations, is uniform, $p_i=1/M$.  The entropy just counts the number of states, $S=k_BT \log_2 M$. The entropies in other ensembles, including the canonical ensemble, are then introduced via a Legendre transform. For example in the canonical ensemble one allows for the energy to fluctuate, keeping the mean energy fixed. As a result, the system can now find itself in many energy states, with different probabilities. Here we will start with directly defining the canonical entropy:
\beq
S=-\sum_i p_i \log_2 p_i,
\eeq
which will be a key quantity of interest. In information theory and computer science the canonical entropy (up to the choice of units) is referred to as Shannon entropy. The intuition behind this form of entropy is similar to that of the microcanonical entropy -- it counts the number of accessible states, but now all of these states need not be equally likely. We are trying to define a measure of ``accesible states,'' but if their probabilities are unequal, some of the states are in fact less accessible than others. To correct for this we must weigh the $\log_2{p_i}$ contribution to the entropy by the the probability $p_i$ of observing that state, $S=-\sum_i p_i \log p_i$. By convention used in information theory we chose the units where $k_BT=1$. The logarithm base 2 defines a unit called a \emph{bit}, which is an entropy of a binary variable that has two equally accessible states.  In general in the case of $M$ equally probable states, we recover
\beq
S=-\sum_{x=1}^M 1/M \log_2 (1/M)=\log_2 M \;\;\mathrm{[bits]}.
\eeq
According to this formula, the uncertainty in the outcome of a fair coin toss is 1 bit, whereas the uncertainty of an outcome with a biased coin is necessarily less than 1 bit, allowing the owner of such a coin to make money in betting games. Entropy is nothing else but a measure of the uncertainty of a random variable distributed according to a given distribution $P=\{p_i\},i=1,\dots,M$. Entropy is always positive, measured in bits, and in the discrete case always takes a value between two limits: $0\leq S[P] \leq \log_2 M$. The entropy (uncertainity) is zero when the distribution has its whole weight of $1$ concentrated at a single $i$. The entropy (uncertainty) is maximal when $p_i=1/M$, i.e. $P$ is a uniform distribution. 

The notion of Shannon entropy generalizes to continuous distributions, and to functions of several variables, such as concentrations of many types of proteins $\vec{c}=\{c_1, c_2, ...,c_M\}$ in a gene regulatory network:
 \beq
S=-\int d\vec{c} \;p(\vec{c})\log_2 p(\vec{c}). \label{Sform}
\eeq

As in thermodynamics, the entropy cannot be measured directly. In physics one often measures specific heat, which is connected to a difference of entropies, to gain insight about the number configurations accessible to the system. To illustrate this, consider a cell with concentration $c$ of proteins that fluctuates around its mean $\bar{c}$ and is well approximated by a Gaussian of width $\sigma_c$:
\beq
P(c)=\frac{1}{\sqrt{2\pi \sigma_c^2}} e^{-\frac{(c-\bar{c})^2}{2\sigma_c^2}}.\label{Gaussian_c}
\eeq
Following Eq~(\ref{Sform}), the entropy of $P(c)$ is $S=\log_2\sqrt{2\pi e \sigma_c^2}$. First, we observe that the entropy does not depend on the mean $\bar{c}$, since the number of accessible states does not depend on where in phase space these states are located. Next we see that, somewhat counterintuitively, the entropy seems to depend on a choice of units: if the units of concentration (and therefore $\sigma_c$) change, the value of the entropy changes as well. This is a reflection of the fact that $c$ is a continuous variable and the (discrete) number of accessible states must depend on how finely we measure small differences in $c$; nominally, if $c$ were known with arbitrary precision, the number of states would be infinite. However, as long as we are only interested in difference of entropies, or if we specify the measurement precision and discretize $c$ by binning, no practical problems arise.\footnote{Formally, these problems are addressed by working with ``differential entropy,'' or Kullback-Leibler distance, instead of entropy, where the entropy of the distribution of interest $P(c)$ is defined only relative to some prior distribution $P_0(c)$ \cite{cover+thomas_91}.} As we will soon show, the information measure that we are pursuing is indeed a difference of entropies, and the issues with continuous distributions will not cause us any problems.

Having discussed entropy as a measure of uncertainty, it is time to return to our original goal of computing how much our uncertainty about the output $g$ is reduced by knowing the value of the input $c$. Let $P(g|c)$ describe the input/output relation in a $c\rightarrow g$ regulatory element. Then the entropy of this conditional distribution will measure the uncertainty in $g$ if we know $c$, that is, it will measure the ``number of accessible states'' in $g$ consistent with the constraint that they happen when input $c$ is presented:
\beq
S[P(g|c)]=-\int dg\; P(g|c)\log_2P(g|c).
 \eeq
Note that this entropy still depends on on the input $c$ (but no longer on $g$, which has been integrated out).

Now suppose for the moment that we did not know the value of the input $c$. In that case the uncertainty about the value $g$ would be directly $S[P(g)]=-\int dg\; P(g)\log_2 P(g)$. If we form a difference of the two entropies, we can measure how much our uncertainty about $g$ has been reduced by knowing $c$:
\beq
\Delta S=S[P(g)]-S[P(g|c)]
\eeq
We can repeatedly measure this entropy difference in different input concentration regimes, and take an average according to the distribution $P(c)$ with which the inputs are presented. The resulting quantity, central to our discussions, is called \emph{mutual information}: 
\beq
I(c;g)=\int dc \; P(c) \left(S[P(g)]-S[P(g|c)] )\right). \label{mut1}
\eeq
Briefly, this quantity in bits measures how much, on average, our uncertainty in one variable (e.g. $g$) has been decreased by knowing the value of a related variable (e.g. $c$). Mutual information is a scalar number (not a function!), and it is customary to write $c$ and $g$ in parenthesis separated by a semicolon as in Eq~(\ref{mut1}) to denote between which two variables the mutual information has been computed.

Using the defintions of the entropies and conditional entropies
\beq
P(g,c)=P(g|c) P(c),
\eeq
we can reformulate the information between the input and output as:
\begin{eqnarray}
I(c;g)&=&\int dc\int dg\; P(c,g)\log_2\frac{P(g,c)}{P(c)P(g)}\\ 
&=&\int dc \; P(c)\int dg\; P(g|c)\log_2\frac{P(g|c)}{P(g)}\\
&=&\int dg\; P(g)\int dg\; P(c|g)\log_2\frac{P(c|g)}{P(c)}. \label{mut2}
\end{eqnarray}
From this we clearly see that information is a symmetric quantity -- the information the input has about the output is the same as the information the output has about the input. Hence this measure of information is called mutual information. We also clearly see that if the joint distribution of inputs and outputs is independent, $P(c,g)=P(c)P(g)$, then $I(c;g)=0$. In this case the entropy of the whole system would be the sum of the individual entropies. If the variables are not independent the entropy of the system is reduced by the mutual information:
\beq
I(c;g)=S[P(c)]+S[P(g)]-S[P(c,g)].
\eeq
Mutual information also has other interesting properties:
\begin{itemize}
\item {\bf It can be defined for continuous or discrete quantities.} Mutual information is a functional of a probability distribution, and probability distributions are very generic objects. $c$ and $g$ could both be continuous, or any one or both can be discrete.
\item {\bf It is reparametrization invariant.} Mutual information betwen $c$ and $g$ is the same than mutual information between any one-to-one function of $c$, $f(c)$, and any one-to-one function of $g$, $h(g)$, that is $I(c;g)=I(f(c);h(g))$. In biological context, this is a great asset: experiments often report, e.g. intensities or log-intensities on the microarray chips or in FACS sorting, and there is a lot of discussion about how this data should be normalized, transformed or interpreted prior to any analysis, or how the cells themselves ``interpret'' their internal concentrations of TFs. This feature of mutual information is important because other statistical measures of correlation, like correlation coefficients, depend on transformations of the data. Mutual information, in contrast, is invariant to such reparametrizations of the variables.
\item {\bf It obeys data processing inequality.} Suppose that $g$ depends on $c$ and $k$ depends on $g$ (but not directly on  $c$), in some probabilistic fashion. In other words, one can imagine that there is a Markov process, $c\rightarrow g \rightarrow k$, where arrows denote a noisy mapping from one value to the next one: $c$ gives rise to $g$ and $g$ to $k$. Then $I(c;k)\leq I(c;g)$, that is, information necessarily either gets lost or stays the same at each noisy step in the transmission process, but it is never ``spontaneously'' created.
\item {\bf It has a clear interpretation.} If there is $I$ bits of mutual information between input $c$ and output $g$, this can be interpreted as there being $2^{I(c;g)}$ \emph{distinguishable} levels of $g$ that can be reached by dialing the value of input, $c$. By ``distinguishable'' we mean distinguishable given the intrinsic noise in the channel $c\rightarrow g$. 
\end{itemize}

There is a number of powerful theorems relating to mutual information which we will not go into here, but the interested reader is referred to the classic text of Thomas and Cover for details \cite{cover+thomas_91}. 

Let us consider an instructive example of a Gaussian channel. We assume the input/output relation between $c$ and $g$ is linear (or nonlinear, but can be linearized around the operating point):
\beq
g=c+\eta,
\eeq
while the noise in this $c\rightarrow g$ process is additive and drawn from a Gaussian distribution:
\beq
P(\eta)=P(g|c)=\frac{1}{\sqrt{2 \pi \sigma^2}} \exp\left(-\frac{(g-c)^2}{2\sigma^2}\right);
\eeq
note that in this simple example the variance is not a function of $c$ [as in our models of gene expression, e.g. in Eq~(\ref{noisefinal})].
Let us assume that the input $c$ itself is also a Gaussian distributed random variable, given by the distribution in Eq~(\ref{Gaussian_c}). Having fixed the distributions of the noise and the input, this uniquely defines the output to be Gaussian as well. Using Eq~(\ref{mut1}) we find the mutual information between the input and output to be:
\beq
I(c;g)=\frac{1}{2}\log_2\left[1+\frac{\sigma_c^2}{\sigma^2}\right],\label{gaussian_bound}
\eeq
where $\sigma_c^2/\sigma^2$ is  the ratio of the signal variance to the noise variance, often referred to as the \emph{signal-to-noise ratio} or SNR.

If, as in our example, the noise is Gaussian and additive, then one can show that the information transmission is maximized at fixed input variance when input is drawn from a Gaussian distribution, as we assumed above \cite{cover+thomas_91}. This is related to the fact that Gaussian distribution is a distribution that maximizes the entropy for a fixed variance, and that information is maximized, according to Eq~(\ref{mut1}), when the output (or input) entropies are maximized. Let us show that the Gaussian distribution really maximizes the entropy subject to a variance constraint. 
We formulate the problem as a constrained optimization procedure, for the input distribution $P(c)$:
\begin{eqnarray}
{\cal{L}}[P(c)]&=-&\int dc\; P(c) \log P(c) -\lambda_0\int dc\;P(c) \nonumber\\
&-&  \lambda_1\int dc\; c P(c)-\lambda_2\int dc\; c^2 P(c).
\end{eqnarray}
Here, the Lagrange multiplier $\lambda_0$ will enforce that the distribution is normalized, $\lambda_1$ can be used to fix the mean, and $\lambda_2$ to constrain the variance. 
Optimizing with respect to $P(c)$, $\delta \mathcal{L}/\delta P(c)=0$, we obtain:
\beq
 \log P(c) = -1 -\lambda_0-   \lambda_1 c-   \lambda_2 c^2.
\eeq
We can complete the square in $c$ and express $P(c)$ to obtain:
 \beq
P(c) = Z^{-1} \exp\left(-\lambda_2\left[c+\frac{\lambda_1}{2\lambda_2}\right]^2\right),
\eeq
where $Z=\exp(-1-\lambda_0-\lambda_1^2/4\lambda_2)$. By making the identifications $\bar{c}=-\lambda_1/2\lambda_2$ and $\sigma^2_c=1/2\lambda_2$, we see that we can select Lagrange multipliers such that the result of the entropy maximizing optimization is a Gaussian distribution with the desired mean and variance.
These arguments together show that for a channel with Gaussian additive noise with fixed variance, mutual information is maximized when the input and output variables are chosen from a Gaussian distribution. The Gaussian channel result of Eq~(\ref{gaussian_bound}) gives an upper bound on the amount of information that can be transmitted under these assumptions. We will return to Gaussian channels when considering time-dependent solutions in Section \ref{timedep}.

Finally, we present the last example, originally studied by Laughlin  in the context of neural coding of contrast in fly vision \cite{Laughlin,spikes}, to build intuition about maximal information solutions. Consider a nonlinear system that translates an input $c$ to an output $g$, via a mean input/output relation $\bar{g}=\bar{g}(c)$. In Laughlin's case, the input was the contrast incident on the fly's eye, while the output was the firing rate of a specific neuron in the fly visual system; the input/output relation in this case was experimentally measurable quantity. The system is stochastic and so we really measure $g$, which is a random variable whose mean is given by $\bar{g}(c)$. Let us assume the noise is additive and constant -- it does not depend on the value of $c$ [this assumption is the main difference between this problem in fly vision and the case of gene regulation which we study below, where both mean input/output relation and the noise are  functions of $c$]. We can then ask, as Laughlin did, what distribution of inputs, $P(c)$, will maximize information transmission through this channel, by writing down an optimization problem for the information of Eq~(\ref{mut1}), while constraining the normalization of $P(c)$:
\beq
\frac{1}{\delta P(c)} \left[ S[P(g)]-\int dc\; P(c) S[P(g|c)] + \lambda \int dc\;P(c)\right]=0,
\eeq
where we are considering $P(g|c)$ as given from the experiment and fixed, and $P(g)=\int dc\;P(c)P(g|c)$.
If noise is independent of $c$, then the conditional entropy is also a constant, $S[P(g|c)]=\alpha$. Optimizing $\cal{L}$ we obtain:
\beq
\left(-\alpha + \lambda \right)+\int dg\; \frac{\delta S[P(g)]}{\delta P(g)} P(g|c)=0
\label{unidist1}
\eeq
The second term gives:
\begin{eqnarray}
&&\int dg\; P(g|c) \frac{\delta S[P(g)]}{\delta P(g)} \\
&=& \int dg\; P(g|c) (-\log P(g)-1) \\
&=&-\log P(\bar{g})-1,
\end{eqnarray}
where in the last line we have assumed $P(g|c)$ is strongly peaked  around the mean, $\bar{g}(c)$ (this enables us to approximate the average over $\log P(g)$ with the log of the distribution of average values). Apart from $\log P(\bar{g})$ all terms in Eq~(\ref{unidist1}) are constant, hence we have derived the result the information-maximizing distribution of mean outputs is a constant as well:
\beq
P(\bar{g})=\mathrm{const}.\label{unidist2}
\eeq
Since $P(c) dc=P(\bar{g}) d\bar{g}$, we find using Eq~(\ref{unidist2}) that
\beq
P(c)=\frac{d \bar{g}(c)}{d c}. \label{laugh}
\eeq
The optimal way to encode inputs, given a known input/output relation $\bar{g}(c)$ and constant noise, is such that all responses $\bar{g}$ are used with the same frequency. In Laughlin's case, this result made a prediction: if the fly visual system is adapted to the distribution of contrast levels in the environment, then by measuring $\bar{g}(c)$ one could predict the distribution of contrast levels in nature according to Eq~(\ref{laugh}). This prediction can be checked by going outdoors and collecting the natural contrast distributions directly using a properly calibrated camera. The results matched the predictions beautifully, illustrating that the fly visual neuron is using its finite dynamic range of firing rates optimally. In engineering, this encoding technique that takes an arbitrary input distribution $P(c)$ and transforms it into a uniform output distribution $P(\bar{g})$, is known as \emph{histogram equalization}.

We emphasize again that this result is only true if the noise is constant, and if the distribution $P(g|c)$ is tightly peaked around the mean value, $\bar{g}(c)$. If these assumptions do not hold, but the range of inputs is constrained, the optimal input distributions may be discrete (a sum of delta functions) \cite{huangmeyn}. In Section \ref{SNA} we shall see how the optimal input and output distributions change when the noise depends explicitly on the input $c$.

In this review we are considering how information is transmitted between the input of a gene regulatory network and its output. Apart from mutual information that we are using,  there exist other measures of information, which ask slightly different questions. For example, Fisher information tells us how well one can estimate (in a L2-norm sense) the value of an unknown parameter $\theta$ that determines the probability distribution from which measurements are drawn. However, just as Fisher information makes assumptions about the ``error metric'' (L2 norm), so do other measures make alternative assumptions either about the distributions from which the data are drawn or about the error metric. Shannon has shown that mutual information alone provides a unique, assumption-free measure of dependency for any choice of $P(c,g)$ \cite{shannon_48}.

\subsection{Information transmission as a measure of network function}\label{Sec_justifyinfo}
In previous sections we laid down the mathematical foundations for describing gene regulation and introduced the concept of information transmission between inputs and outputs of noisy channels. Before bringing the two topics together and showing how information transmission applies to genetic regulatory networks,  we should ask ourselves why information transmission  might even be a feasible  measure of network function.  In this section we briefly review some experimental justifications for this approach.

The main criticism against any given (mathematically definable or tractable) measure of function, including information, is the lack of arguments why this particular measure should be singled out from other candidate measures. In words it is clear  that ``selection is acting on the function,'' but the biological function in this context is often thought to be some arbitrarily complicated mathematical function that could weight many aspects of the network together in some uncomprehensible way. Precisely for this reason we use mutual information: regardless of what exactly the biological function is and how the network processes the inputs, according to Shannon, there \emph{has} to be some minimal amount of transmitted information to support this biological function. 

 A stronger criticism states that he examples of networks in extant species observable today are not yet optimized for biological function, whatever that might be. If they are not even close to the extremal point and the space of the networks is large, then the observable networks today could be viewed  purely as results of their ancestry, as random draws from a huge space of possible networks that perform the biological function just ``well enough'' for the organisms to survive. This is certainly a valid criticism, but it is hard to see what one can do about it \emph{a priori}. However, if it turns out that the networks observed today are at (or close to) the extremum of some measure of function that we postulate, such assumptions might be validated \emph{a posteriori}. While valid, this criticism should therefore not prevent us from trying to find relevant network design principles.

On the other hand, networks do have to obey physical laws and constraints, such as the limitations in accuracy of \emph{any} network function due to stochasticity in gene regulation. It is therefore interesting to explore how these limits translate into observable circuit properties. There could be other constraints shaping the network structure apart from noise: the metabolic cost to the number of signaling molecules used by the network, or the constraint on the speed of signaling etc. We decided to concentrate on the noise constraint (which is indeed related to the constraint on the number of signaling molecules, as we will show later) because it has physical basis relevant for all networks, and because it can be measured in today's experiments. 

Taken together, we realize that not all (if any) gene regulatory networks are solely optimized to transmit information. However, as we have argued above, mutual information is in some sense a minimal measure: any network that performs whichever biological function well will have to keep noise in check, and better performance of that function will imply smaller noise and thus larger values of transmitted information. In this sense our approach can fail if constraints other than noise are dominant: then information will fail to discriminate between good networks (in information sense) that nevertheless differ strongly in terms of these remaining constraints.

As we show below, the principle of maximizing information in genetic networks is predictive about network structure. Therefore, theoretical results can be compared to experiments, which in turn can give us insight into other principles and constraints at play in nature. In the long run we are thus hoping for a productive interaction between theory and experiment that systematically reveals various determinants of genetic regulatory networks.

One of the systems in which ideas about information transmission in genetic regulatory networks could be tested has been early embryonic development of \emph{Drosophila}. This genetic organism is a prime example of  spatial patterning, where nuclei in the early embryo, though they all share the same DNA, initiate different programs of gene expression based on a small number of maternal chemical cues. These precise and reproducible spatial domains of differential gene expression in the embryo that later lead to patches of cells with distinct developmental fates have been extensively studied, as has been the nature of the maternal cues, called \emph{maternal morphogens}. In genetics and molecular biology researchers have thus introduced already the concept ``positional information'' encoded in the maternal morphogens, which is read out by the developmental regulatory network, but this concept has not been defined mathematically. In the following paragraphs we will very briefly outline the biology of early \emph{Drosophila} development, review the relevant measurements, and proceed to connect them to the framework we built in the preceding sections.

\begin{figure}
\includegraphics[width =  \linewidth]{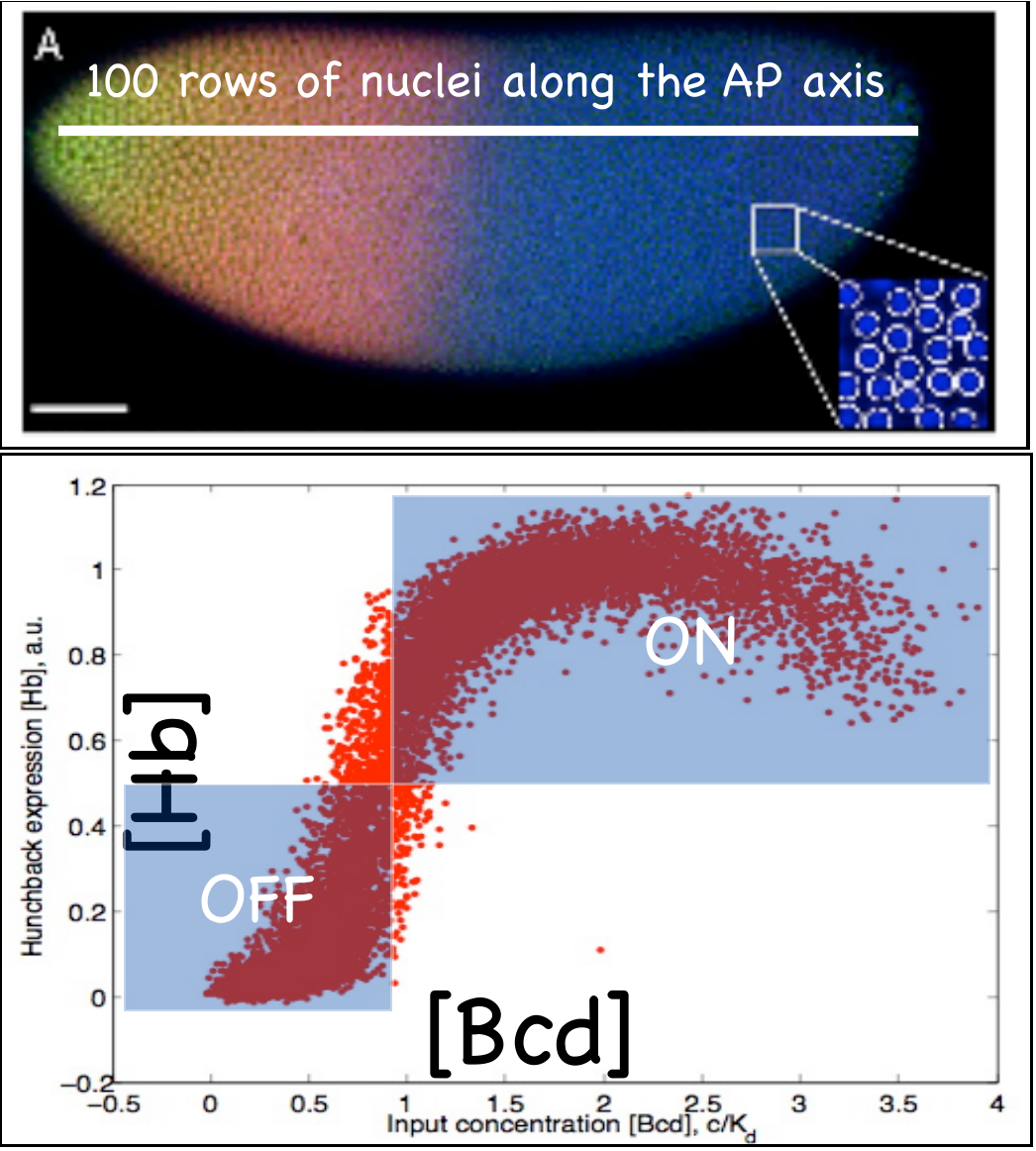}
\caption{\emph{Drosophila melanogaster} embryo at cell cycle 14. Nuclei stained in blue (see inset), bicoid stained in green and hunchback stained in red, data reproduced from Ref \cite{Gregor}. At this stage, about 6000 nuclei are present in the embryo, of which about a quarter are visible under a single microscope view. Each nucleus provides a joint quantitative readout proportional to bicoid and hunchback intensities; the data is shown in scatter plot below. Usually hunchback was understood as having a single precise boundary that separates the domain of high expression (``on'') from the domain of low expression (``off''). We  use information theory to make this statement precise and to find out if the bicoid/hunchback regulatory element really can be understood just as a binary switch. }
\label{f-droso-ap}
\end{figure}

When a \emph{Drosophila} egg is produced by the mother, the mother deposits mRNA of a gene called Bicoid  in the anterior portion of the egg. These mRNAs are translated into into bicoid protein, which diffuses towards the posterior, establishing a decaying anterior--posterior protein gradient (see Fig \ref{f-droso-ap}). The maternal morphogen bicoid acts as a transcription factor for four downstream genes, known as ``gap genes" (Hunchback, Kr\"uppel, Knirps and Giant). Looking along the long axis of the ellipsoidal egg, known as the AP (anterior-posterior) axis, one can see about 100 rows of nuclei at cell cycle 14, about 2 hours after egg deposition, when the nuclei still uniformly tile the surface of the egg and before large morphological rearrangements, called \emph{gastrulation}, start to occur. These nuclei express  proteins (mostly transcription factors) that will confer cell fate: nuclei belonging to various spatial domains of the embryo express specific combinations of genes that will lead these nuclei to become precursors of different tissues. Stainings for relevant transcription factors  have shown a  remarkable degree of precision with which the spatial domain boundaries are drawn in each single embryo, and a stunning reproducibility in positioning of these domains between embryos. Although probably a slight overgeneralization, we can say that at the end of cell cycle 14, along the AP axis, each row of nuclei reliably and reproducibly expresses a gene expression pattern that is characteristic of that row only -- in other words, the nuclei have unique \emph{identities} encoded by expression levels of developmental TFs along the long axis of the embryo.

The spatial gradients form a chemical coordinate system: it is thought that each nucleus can read off the local concentration of bicoid (and other morphogens), and based on these inputs, drive the expression of the second layer of developmental genes (the gap genes, which we denote by $g_i$); these in turn lead to ever more refined spatial patterns of gene expression that ultimately generate the cell fate specification precise to a single-nuclear row. For a recent review of the gap gene network, please see Ref~\cite{jaeger2011}.

We can make a simple back-of-the-envelope calculation: If there are 100 distinguishable states of gene expression along the AP axis responsible for 100 distinct rows of nuclei, some mechanism must have delivered $I\approx \log_2(100)\approx 7$ bits of information to the nuclei. That's the minimum amount of information needed to make a decision about the cell fate along the AP axis. Intuitively, this number is the same as the minimum of how many successive binary (``yes or no'') questions are needed to uniquely identify one item out of 100: the best strategy is to ask such that each question halves the number of options remaining.  Each answer to the question would thus convey 1 bit of information, and reduce the initial uncertainty of 7 bits by 1 bit. Similar patterning mechanisms also act along the other axes of the embryo, and if each of the 6000 nuclei at cell cycle 14 were uniquely determined, these systems together would have to deliver about 13 bits of information.

Let us start by considering the regulation of Hunchback by bicoid. By simultaneously observing the concentrations of bicoid ($c$) and hunchback ($g$) across the nuclei of an embryo, one can sample the joint distribution $P(c,g)$, see Fig~\ref{f-droso-ap}. Usually it was assumed that hunchback provides a sharp, step-like response to its input, bicoid; mathematically, this would mean that the bcd/hb input/output relation is switch-like, with an ``on'' and an ``off'' state, yielding information transmission capacities of about 1 bit. However, is this really the case?

Using the methods from Section~\ref{info} combined with the direct experimental measurements of probability distributions of Gregor and collaborators \cite{Gregor}, one can find how much information bicoid $c$ and hunchback $g$ carry about each other. The result is $I_{\rm expt}(c;g)=1.5\pm 0.1$ bits, where the error bar is computed across 9 embryos. This is an experimentally determined quantity, and the errors [apart from the estimation bias \cite{slonimetal}] are related mostly to our ability to fairly sample the distribution $P(c,g)$ across the ensemble of nuclei. Our sampling is not complete because a single microscope view only records about a quarter of all nuclei, but we believe that that sampling is not very biased. Another point to have in mind is that the computation of $I(c;g)$ reflects all statistical dependency in the probabilistic relation $c\rightarrow g$: both the direct regulation, as well as any possible indirect regulation through an unknown intermediary $x$, e.g. $c\rightarrow x \rightarrow g$. Thus, for example, if bicoid activates hunchback which self-activates itself, our information estimation has taken this into account. If, however, $g$ is regulated also by an input $y$ independent of $c$, that is $\{c,y\}\rightarrow g$, and our experiment does not record $y$, then we might be assigning some variability (or noise) to $g$, although that noise really would be a systematic regulatory effect caused by $y$. In this last case, we would measure a smaller value of $I(c;g)$  and would underestimate the real precision in the system; the true value would only be revealed upon recording the unobserved regulator $y$ and computing $I(\{c,y\};g)$. This might be the case for bicoid regulating Hunchback, since we know that {\bf (i)} some hunchback is also maternally deposited (not all hunchback is made under control of bicoid); {\bf (ii)} nanos, another maternally supplied mRNA, establishes a separate protein gradient extending inwards from the posterior, and inhibits the translation of Hunchback; {\bf (iii)} there might be weaker influences from other morphogens and terminal patterning factors.

Having these caveats in mind, our first finding is that the information transmission of 1.5 bits between bicoid and hunchback that we measure from the data is larger than 1 bit, which would be needed if bicoid/hunchback transformation were a simple binary switch. To our knowledge this was one of the first times that a quantitative measure of ``regulatory power'' was computed for a genetic regulatory element that was measured in a high-precision experiment. 

\begin{figure}
\includegraphics[width =  \linewidth]{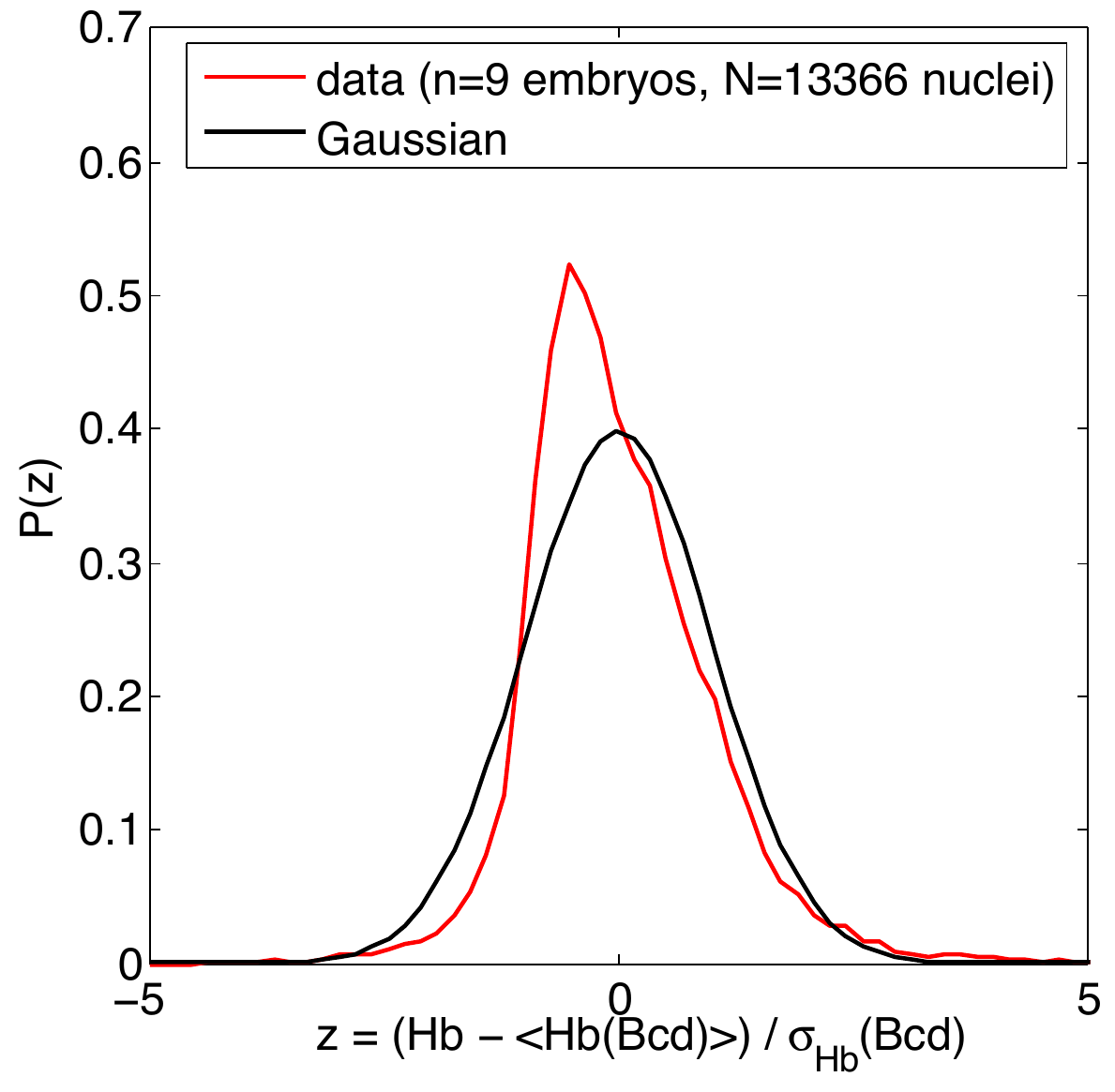}
\caption{The noise in the regulation of hunchback is approximately Gaussian. Joint nuclear measurements of bicoid and hunchback are performed across $\sim$13k nuclei in 9 \emph{Drosophila} embryos at nuclear cycle 14; data from Ref~\cite{Gregor}. Nuclei are sorted in 100 bins according to their bicoid concentration; for each bin, we compute the mean hunchback response, $\langle\mathrm{Hb}(\mathrm{Bcd})\rangle$, and the noise in the response $\sigma_{\mathrm{Hb}}(\mathrm{Bcd})$. For each nucleus we take its input bicoid concentration, find the mean response and noise for that bicoid level and define its z score as the deviation from the mean, normalized to the noise. The plot shows a distribution of the z scores across all nuclei (in red) and compares it to the case where the noise would be perfectly Gaussian (black) with zero mean and unit standard deviation. The agreement is reasonable, with real data being somewhat more skewed.}
\label{f-gaussnoise}
\end{figure}

While the result that 1.5 bits estimated from the data is larger than 1 bit needed for a binary switch is intriguing, it would be instructive to have another measure to compare 1.5 bits to. To this end, we will put an upper bound of how much information could have maximally been transferred between bicoid and hunchback, given the measured level of noise in the system. To do this, let us start by writing:
\begin{equation}
P(c,g)=P(g|c)P_{TF}(c).
\end{equation}
As shown in Section~\ref{Sec_molmodels}, the term $P(g|c)$ describes the input/output properties of the regulatory element. From experiment, we can determine the mean response $\bar{g}(c)$ of the regulatory element and the noise in the response, $\sigma_g^2(c)$. In Fig~\ref{f-gaussnoise} we show that the noise found directly from the measurements, $p(g|c)$, is to a good approximation Gaussian $\mathcal{G}$. Therefore these two measurements, $\bar{g}(c)$ and $\sigma_g^2(c)$, determine $P(g|c)$ to a good approximation.

%insert figure here

To ask about the maximum achievable information transmission given the measured input/output relation $P(g|c)\sim\mathcal{G}(g;\bar{g}(c),\sigma_g(c))$, we proceed in a manner similar to that used by Laughlin in his studies of fly vision. We write the Lagrangian
\begin{equation}
\mathcal{L}[P_{TF}(c)]= I(c;g) - \Lambda\int dc\;P_{TF}(c),
\end{equation}
where $\Lambda$ is a Lagrange multiplier that will enforce the normalization of $P_{TF}(c)$, while 
\begin{equation}
I(c;g)=\int dc\;P_{TF}(c)\int dg\; P(g|c) \log_2\frac{P(g|c)}{P(g)}
\end{equation}
is the mutual information, and $P(g)=\int dc\;P_{TF}(c)P(g|c)$. We can now look for the optimal distribution of inputs, $P_{TF}(c)$, which must satisfy:
\begin{equation}
\frac{\delta\mathcal{L}[P_{TF}(c)]}{\delta P_{TF}(c)}=0. \label{variation}
\end{equation}
One way to solve this variational problem is numerically. For details see Refs \cite{ggpnas,ggpre,blahut}; here we only report on the results. 

\begin{figure}
\includegraphics[width =  \linewidth]{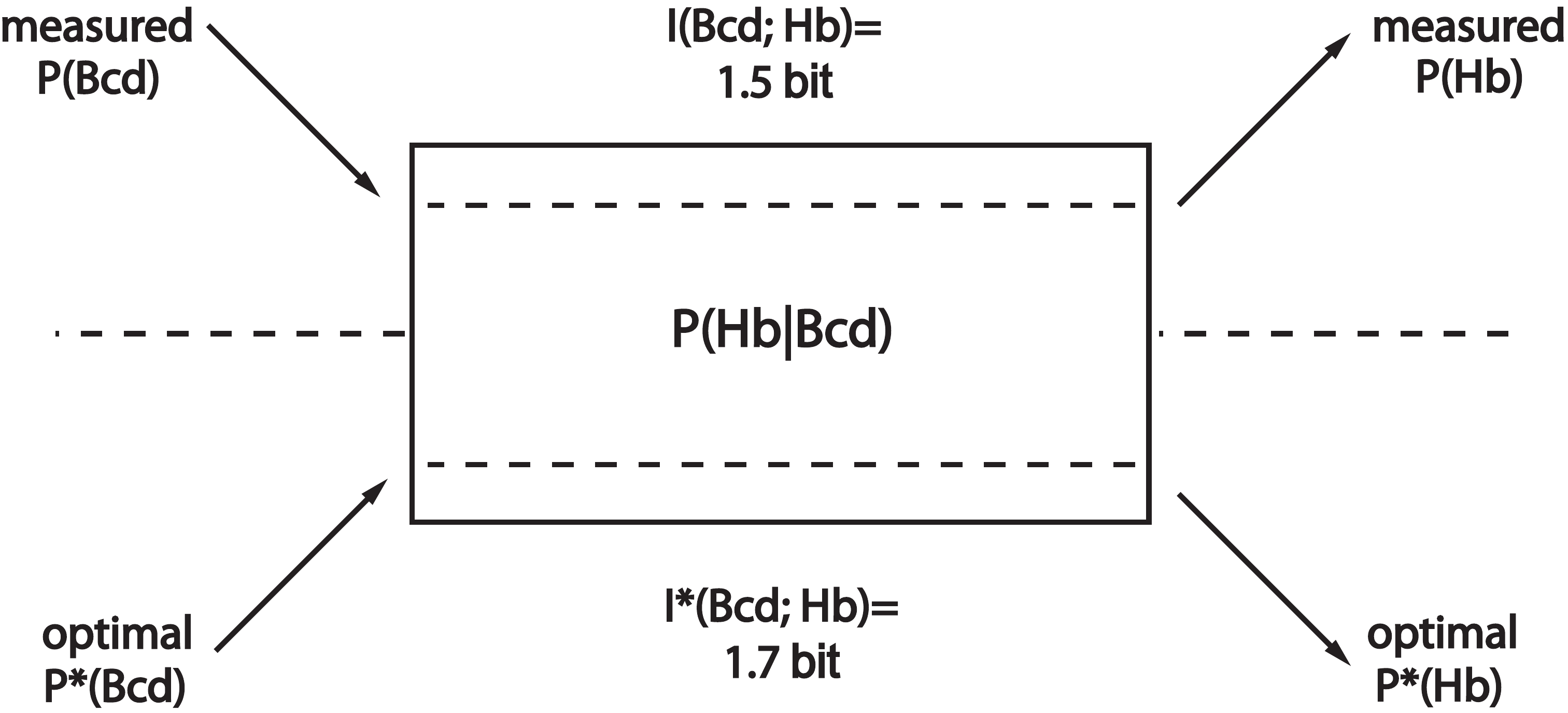}
\caption{The real (measured) information transmission and the maximal information transmission (channel capacity) in the bicoid/hunchback regulatory system. The input/output relation $P(g|c)=P(\mathrm{Hb}|\mathrm{Bcd})$ is measured and held fixed. To estimate the true information transmission of 1.5 bits, the experimentally sampled $P_{TF}(\mathrm{Bcd})$ is used to construct the joint $P(c,g)$. To find the channel capacity, $P_{TF}(\mathrm{Bcd})$ is varied until the information-maximizing choice is found numerically, denoted as $P^*(\mathrm{Bcd})$; this yields 1.7 bits of capacity. The optimal choice for the input distribution also predicts the optimal distribution of outputs, shown in Fig.~\ref{f-optinfo}.}
\label{f-infoinfo}
\end{figure}

\begin{figure}
\includegraphics[width =  \linewidth]{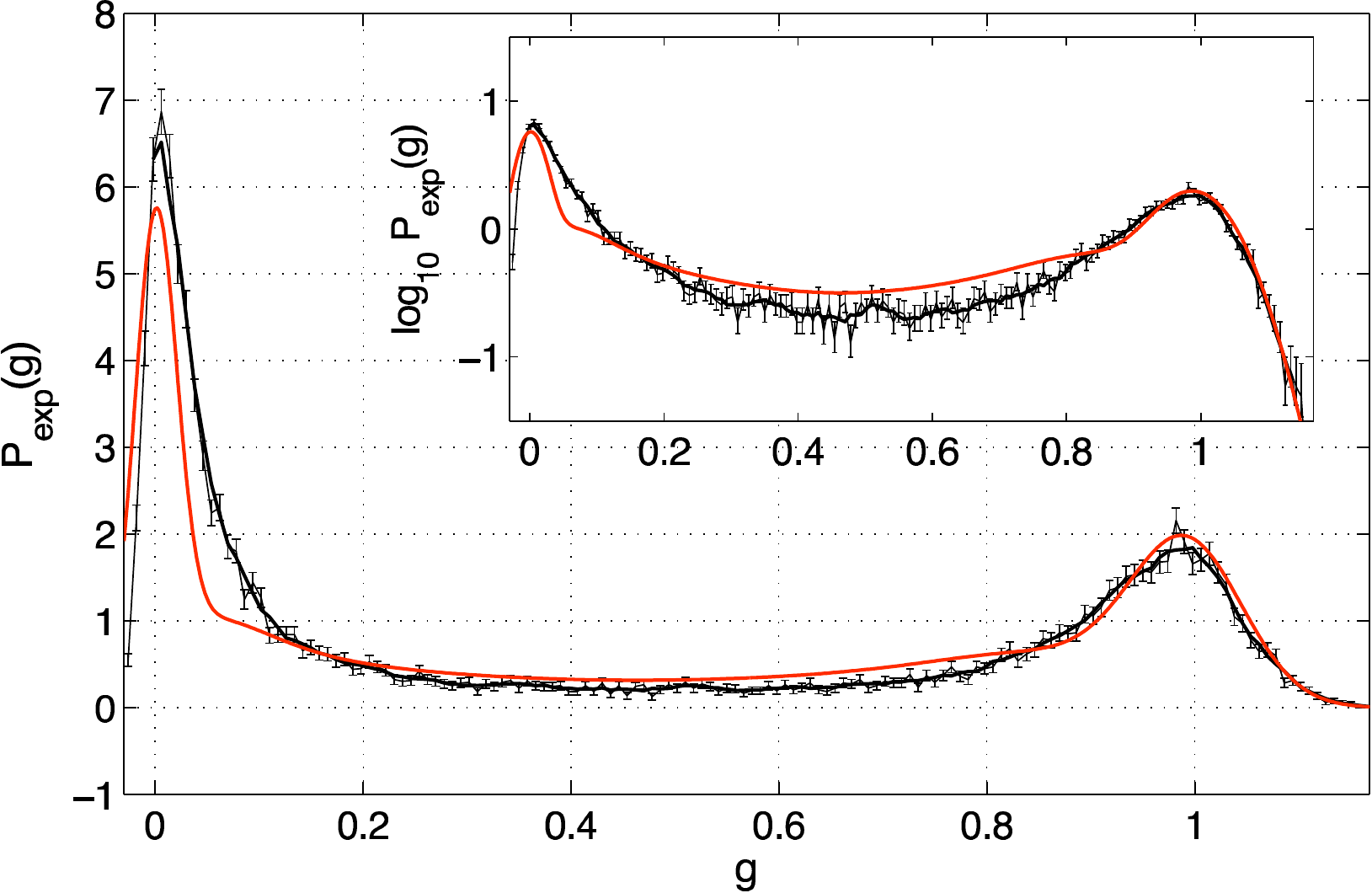}
\caption{The measured (black) and predicted optimal (red) distribution $P(g)$ of hunchback expression levels across an ensemble of nuclei in the \emph{Drosophila} embryo. The expression level $g$ goes from 0 (no induction, posterior) to 1 (full induction, anterior). A considerable fraction ($\sim 30\%$) of nuclei express intermediate levels of hunchback, and the noise in the system is low enough that this intermediate expression level could constitute a separate signaling level from 0 and 1; this would be consistent with the observed information of $1.5$ bits that intuitively corresponds to $2^{1.5}\sim 3$ distinguishable levels of gene expression. The inset shows the same plot on the logarithmic scale.}
\label{f-optinfo}
\end{figure}

We find that holding $P(g|c)$ fixed as determined from the data on bicoid/hunchback relationship, and optimizing $P_{TF}(c)$ numerically, yielded the maximal channel capacity of $I^*(c;g)=1.7$ bits, see Fig.~\ref{f-infoinfo}. Additionally the optimal $P_{TF}^*(c)$ predicts the optimal distribution of hunchback expression levels observed across the ensemble of nuclei, through $P^*(g)=\int dc\; P(g|c) P^*_{TF}(c)$, and the optimally predicted distribution matches the measured distribution very well [Fig.~\ref{f-optinfo}].
The value found for the maximal information transmission (channel capacity) shows that the real biological system is operating close to what is achievable given the noise, that is $I_{\rm expt}(c;g)/I^*(c;g)\approx 90\%$. The high value is somewhat unexpected given that we know that hunchback is regulated also by other inputs, and that bicoid also regulates other targets. Nevertheless this finding is a good motivation to consider taking  maximization of information transmission seriously as a possible design principle.

How should we understand the values in the range of $I\sim 1.5-1.7$ bits? It turns out that the bicoid gradient is read out directly by 4 gap genes: hunchback, kruppel, giant and knirps. If each would independently be able to encode  $\sim 1.5$ bits, then together this genes could convey $I(c;\{g_i\})\sim 6-7$ bits of information about bicoid and would thus achieve the amount needed for AP patterning. In this case, we would be able to claim consistency with the back-of-the-envelope calculation that \emph{requires} at least this amount of information for the AP specification. Before reaching such a conclusion, however, we need to resolve the following issues: {\bf (i)} The readout (gap) genes $\{g_i\}$ are probably not independent, but have some  redundancy, which will mean that they convey less than the sum of their individual information values about $c$; such redundancy, as we find below, can be alleviated by proper network wiring; {\bf (ii)} The next layer of developmental cascade after the gap genes is not regulated \emph{solely} through the gap genes, but receives inputs from maternal morphogens directly; therefore, the gap genes are not a single bottleneck through which the information can flow; {\bf (iii)} especially at the poles of the embryo, gradients other than bicoid provide spatial information about the AP position; {\bf (iv)} our formulation of the problem assumes steady state gap gene readout from a stable gradient; it is not clear that such steady state is really reached in the timeframe necessary for nuclear specification. 

In Section~\ref{info} we briefly described the Gaussian channel approximation, where in addition to Gaussian additive noise one assumes that the input distribution $P_{TF}(c)$ is well-approximated by a Gaussian, and the input/output relation is linear. Clearly, this is not the case at hand: the input/output relation is nonlinear [Fig~\ref{f-droso-ap}], the resulting distributions of hunchback are strongly bimodal [Fig~\ref{f-optinfo}] and the input distribution of Bcd is also not Gaussian [not shown].

%insert figure here

In Ref~\cite{EldonEmberlypre08} Emberly showed that there is an optimal morphogen decay length that minimizes the amount of input proteins that need to be produced, while allowing the target output gene to reach the desired precision. The predictions applied to the bcd/hb system showed that the predicted decay length scale is consistent with the properties of the experimentally observed bcd gradient. Interestingly, Emberly showed that the optimal input bcd gradient also achieves a near maximal transmission of information, making it consistent with the predictions summarized above \cite{ggpnas}.

Further experiments and theory will be needed to successfully address outstanding issues and to check whether the near-optimality in information transmission is maintained as larger portions of the network are recorded experimentally. We hope that the discussion nevertheless provides enough motivation for looking at quantities like $I(x;c)$ -- the information that the morphogen gradient encodes about the physical location $x$; at $I(c;\{g_i\})$, and at $I(x;\{g_i\})$ -- the information that later developmental genes (like gap genes) carry about the physical location. Information processing inequalities also constrain the relationships between these (directly measurable) quantities, providing an implicit check of whether we have missed some unobserved regulatory pathway. Before proceeding, we note that experiments that probe these quantities are not easy, because they require us to measure simultaneously the expression levels of a number of genes, nucleus by nucleus, in order to estimate both the mean response, $\bar{g}_i(c)=\langle g_i(c)\rangle$, as well as the noise covariance in the responses, $C_{ij}(c)=\langle \delta g_i(c)\delta g_j(c)\rangle$, where $\delta g_i = g_i - \bar{g}_i(c)$ and  brackets denote averaging with respect to the ensemble of nuclei.

\section{Information transmission in regulatory networks}
\subsection{Small noise approximation}\label{SNA}

Having seen that in at least one biological system the information transmission approach the channel capacity (maximum achievable transmission given noise), we would like to elevate this finding to a principle: let us find network wiring diagrams and interaction parameters that transmit the most information from input TFs to the regulated output genes. Before we start we should note that this is a very ambitious goal: we are trying to derive (not fit!) the structure of a genetic regulatory network. With all the approximations and simplifications that need to be made (also in the absence of experimentally measured parameters like protein decay times, diffusion constants etc) our standard for success will be if we will have managed to qualitatively reproduce gap gene expression patterns observable in the fly.

Analytically, the problem of finding the maximum information transmission  [Eq~(\ref{variation})] is tractable in the so-called small-noise limit, where across most of the input range the noise over the mean is small, $\sigma_g(c)/\bar{g}(c)\ll 1$. This is the limit which we present and use in the following section  to explore the optimal architecture of small regulatory networks.

We will consider networks where a single transcription factor at concentration $c$ can regulate a set of $K$ target genes $\{g_i\}, i=1,\dots,K$, which may be interacting in a feed-forward network. For now, we will not consider feedback loops that can cause multistable behavior. It is clear that without any constraint, the information transmission can trivially be increased by decreasing the noise, and in biochemical networks noise can be decreased arbitrarily by increasing the number of signaling molecules, both on the input side ($c$) and on the output side ($\{g_i\}$). The crucial idea is therefore \emph{to optimize information subject to biophysical constraints, i.e. subject to using a fixed number of signaling molecules}.

With these assumptions in mind, we sketch the derivation of information transmission in the following text; for details see Refs \cite{ggpre,twb, wtb_pre10}. For additional work on information transmission in biochemical networks see Refs \cite{zivetal, tostevintenwoldeprl, wmw}.

The dynamics of gene expression for genes $\{g_i\}$ is given by a generalization of Eq~(\ref{prot1}) which we used for the case of a single gene:
\begin{equation}
\tau\frac{dg_i}{dt}=f_i(c;\{g_j\}) - g_i + \xi_i,\label{langeqnseclast}
\end{equation}
where $\tau$ is the protein lifetime, $\xi_i$ is the Langevin noise force with $\av{\xi_i(t)\xi_i(t')}=\delta(t-t') N_{ij}=\delta(t-t') \delta_{ij} N_i$. Before proceeding we note that in physical units the input $c$ goes between 0 and $c_{\rm max}$, but when we write down the noise strength (again a sum of input and output noise contributions as in the case of a single gene), we note that this problem has a ``natural'' concentration unit, $c_0=N_{\rm max}/D a \tau$, i.e. the maximum number of independent molecules of the output $N_{\rm max}$, divided by the relevant diffusion constant, typical size of the binding site $a$ and the integration protein lifetime $\tau$. This is simply the scale of output noise divided by the scale of the input noise. With this unit in hand we can make the concentrations dimensionless, so that $c\in[0,C]$, where $C=c_{\rm max}/c_0$, and all $g_i\in[0,1]$ as before.

For completeness, we provide the expression for the noise magnitude $N_i$, which is a generalization of the term explained in Section~\ref{Dernoise1}:
\begin{eqnarray}
N_i&=&\frac{\tau}{N_{\rm max}} \big[\bar{g}_i (c)+c \left(\frac{\partial f_i(c;\{g_l\})}{\partial c}\right)^2+\nonumber \\
&&\frac{1}{c_{\rm max}} \sum_k g_k\left(\frac{\partial f_i(c;\{g_l\})}{\partial c}\right)^2 \big]|_{\{g_k=\bar{g}_k(c)\}}; \label{noisen}
\end{eqnarray}
the first term again corresponds to the output noise, and second and third terms in the parenthesis correspond to the diffusion noise (due to the diffusion of $c$ and of other TFs $\{g_l\}$, respectively). 

In Eq~(\ref{langeqnseclast}), $f(c,\{g_j\})\in [0,1]$ is the regulatory (input/output) function, describing the activation rate of gene $g_i$, given the input $c$ and the expression levels of all the other genes. Various regulation functions were discussed in Section~\ref{Sec_molmodels}; for combinatorial regulation, the most flexible one that we have examined was the Monod-Wyman-Changeaux (MWC) regulation function: 
\begin{eqnarray}
f_i(c;\{g_j\})& =& \frac{1}{1+e^{F_i(c,\{g_j\})}},\nonumber \\
F_i(c,\{g_j\})&=& -n^i_c \log(1+c/K^i_c)- \nonumber \\
&-&\sum_j n^i_j \log(1+g_j/K^i_j) + \tilde{L}^i. \label{lgvn}
\end{eqnarray}
In this model, the regulation of $g_i$ is jointly affected by the input $c$ and the level of other gap genes, $\{g_j\}$, which is reflected by the various contributions to $F_i$: every regulatory input to $g_i$ contributes a term to the ``free energy'' $F$, and each such term is parametrized by $n^i_j$, the number of binding site for $g_j$ in the promoter of $g_i$, and $K_j^i$, related to the energy of binding to that binding site; as before, $\tilde{L}$ is the free energy offset between the ``on'' and ``off'' states when no transcription factor is bound. If we want to avoid feedback and multistability, we can always renumber the genes such that each gene $g_i$ only depends on the input $c$ and other genes $g_j$ where $j<i$.

The regulation in a network of a single input $c$ and $K$ target genes $g_i$ is then described by unknown constants $\{\tilde{L}^i, K^i_j, n^i_j, n^i_c, K^i_c\}$. When $n^i_j\rightarrow 0$, the regulation of gene $j$ by gene $i$ is absent, that is, in the wiring diagram the arrow from $g_j$ to $g_i$ disappears.

Before proceeding, we need also to compute the noise in this regulatory network. The noise in $g_i$ is given by two contributions: the output noise from generating a finite number of proteins of $g_i$, and the input diffusive noise because $g_i$ is regulated by $c$ and other $g_j$. The noise in our setup with $K$ target genes is fully determined by a $K\times K$ covariance matrix:
\begin{equation}
C_{ij}(c)=\langle(g_i - \bar{g}_i(c))(g_j-\bar{g}_j(c))\rangle,
\end{equation}
which can be computed from Eqs~(\ref{lgvn}), as shown in Refs~\cite{ggpre,wtb_pre10}. Here we briefly outline how to do this. By linearizing the dynamical equations in Fourier space for the output concentrations in Eqs~(\ref{langeqnseclast}), one obtains a matrix equation of the form (tildes denote Fourier transforms):
\beq
\hat{A}(\omega) \delta {\bf \tilde{g}} (\omega)=\tilde{\xi} (\omega).
\eeq
In a manner completely analogous to Eq~(\ref{ngg}) but generalized to $K$ output genes, we can then compute the  elements of the covariance matrix in Fourier space as:
\beq
 C_{ij}^{-1}=\int \frac{d \omega}{2 \pi} \left[ \hat{A}^{-1}(\omega) \hat{N}\left[\hat{A}^{-1}(\omega)\right]^{\dagger}\right]_{ij},
\eeq
where the noise magnitudes $N_i$ are given by Eq~(\ref{noisen}).

In addition to computing this matrix, we find that there is a \emph{single dimensionless parameter $C$} in our problem, describing the dynamic range of the input, $c\in [0,C]$. This parameter will control the shape of the optimal solutions\footnote{This is true if all genes $\{g_j\}$ have the same parameters (such as diffusion constant and degradation times), an approximation that we make.}. $C$ is the maximal concentration for the input $c$, expressed in ``natural units of concentration,'' which describes the balance between the input and output noise strengths. Large values for $C$ mean that the output noise is dominant over the input noise, while a small dynamic range and therefore small $C$ means that the input diffusive noise in $c$ is the dominant noise in the system. Alternatively, changing $C$ reflects how many input molecules are at the disposal for communication -- larger values of $C$ are more ``costly'' in metabolic terms, but allow more information to be transmitted.

With the noise covariance matrix in hand, the distribution of outputs given the input $c$ is a multivariate Gaussian:
\begin{equation}
P(\{g_j\}|c)=\frac{e^{-\frac{1}{2}\sum_{i,j=1}^K (g_i-\bar{g}_i(c)) C_{ij}^{-1}(g_j-\bar{g}_j(c))}}{(2\pi)^{K/2} \sqrt{|C|}}. \label{gnoise}
\end{equation}
Suppose that we now  ask the opposite question: having seen the values of gap genes $\{g_i\}$, what is the most likely value of $c$ that produced them, and what is the variance in $c$? If the noise is small, $P(c|\{g_j\})$ will also be Gaussian, which can be found from Eq~(\ref{gnoise}) and the Bayes' theorem:
\begin{equation}
P(c|\{g_j\})\propto e^{-\frac{1}{2}\frac{(c-c^*(\{g_j\}))^2}{\sigma_c^2(\{g_j\})}},
\end{equation}
where $c^*(\{g_j\})$ is the most likely value for $c$ that gives rise to the observed $\{g_j\}$, and 
\begin{equation}
\frac{1}{\sigma_c^2(c)}=\sum_{ij}\frac{d\bar{g}_i}{dc} C^{-1}_{ij}\frac{d\bar{g}_j}{dc};
\end{equation}
$\sigma_c$ is the effective noise level in the input that accounts for all the noise in the system\footnote{In small noise approximation one can reassign the noise from the input to the output and vice versa through the mean input/output relation, as shown in Fig.~\ref{f-noiseprop}. }; this effective noise is computable from the noise covariance matrix and the mean input/output relations.
%
%insert figure here of p(g|c) from experiments
%

Following Eq~(\ref{mut1}), the information between the input and the outputs $I(c;\{g_j\})$ is 
\begin{eqnarray}
I(c;\{g_j\}) &=& S[P_{TF}(c)] - \langle S[P(c|\{g_j\})]\rangle_{P_{TF}(c)},
\end{eqnarray}
where the distribution of inputs, $P_{TF}(c)$ is unknown. We want to find the maximal information transmission given the known noise, therefore, we look for the maximum of $I$ with respect to $P_{TF}(c)$, just as we did in Eq~(\ref{variation}), while insisting that $P_{TF}(c)$ be normalized. Following the derivation in Refs~\cite{ggpre,twb} we find that
\begin{equation}
P^*_{TF}(c)=\frac{1}{Z}\frac{1}{\sigma_c(c)},\label{optddist}
\end{equation}
that is, the system should optimally use those input levels $c$ more frequently that have proportionately smaller effective noise. Using this optimal choice the information, in bits, will be:
\begin{equation}
I(c;\{g_j\})=\log_2\frac{Z}{\sqrt{2\pi e}}, \label{infofinal}
\end{equation} 
where $Z=\int dc\; \sigma_c^{-1}(c)$ is the normalization of the distribution in Eq~(\ref{optddist}).

This is as far as we can push analytically; $I(c;\{g_j\})$ still depends through $Z$ on the parameters $\{\tilde{L}^i, K^i_j, n^i_j, n^i_c, K^i_c\}$ that determine the wiring diagram of the network and the strengths of the regulatory arrows. The last remaining task is, therefore, to numerically optimize Eq~(\ref{infofinal}) with respect to these parameters, and examine the structure of optimal solutions.

\subsection{Optimal network architectures}\label{ONA}
We can finally ask what are the optimal input/output relations for $K$ genes $\{g_i\}$, regulated by the single input $c$, if we do or do not allow for mutual interactions between the outputs. These results are a function of $C$, the dynamic range of the input, which is the single parameter of our optimization problem.

\begin{figure}
\includegraphics[width =  \linewidth]{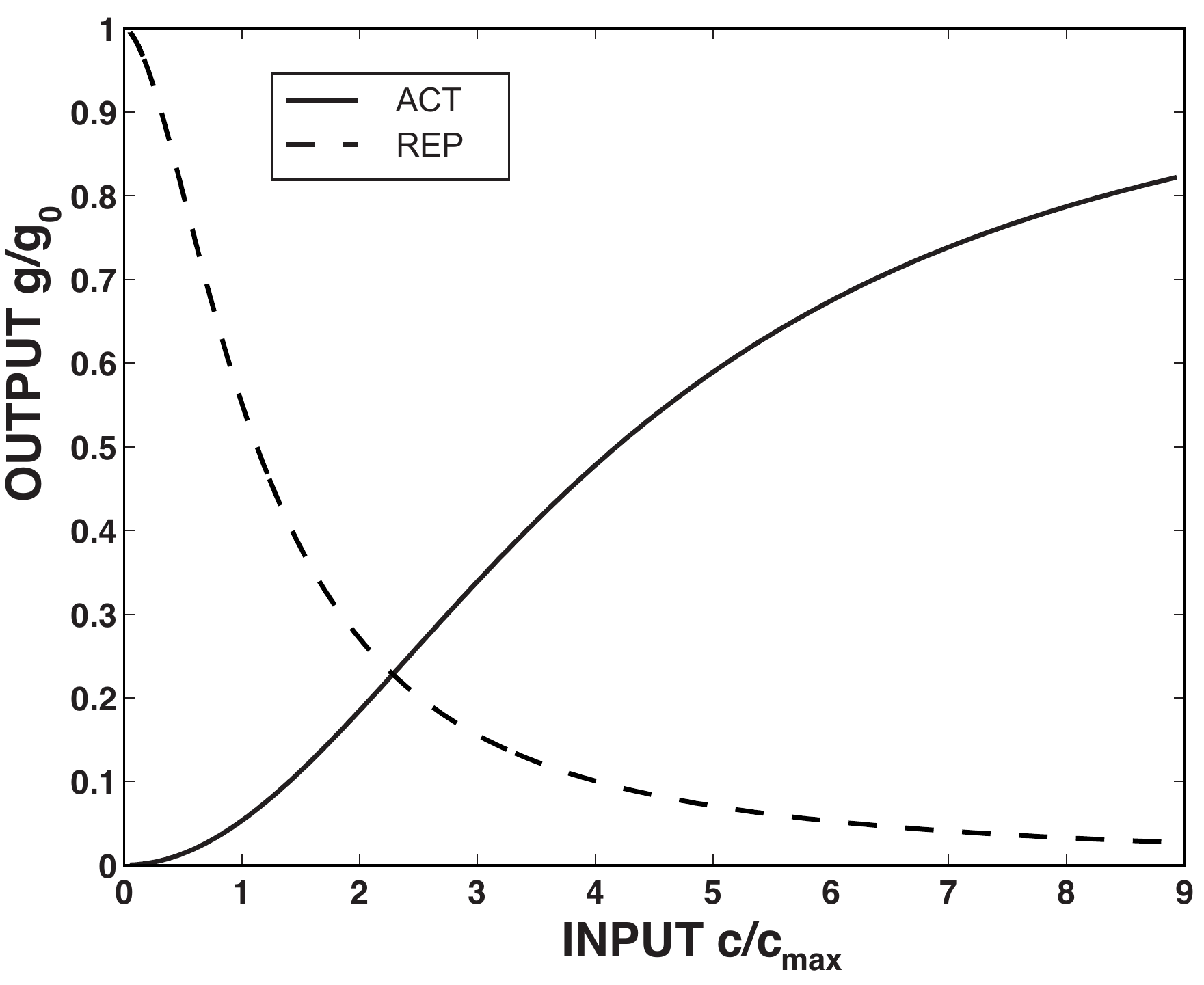}
\caption{The optimal input/output relations for repressors (blue line, dashed) and activators (red line, solid) for one gene $g$ regulated by one input $c$ with no feedback.}
\label{f-oneinput}
\end{figure}

Let us start with considering the simple case of one input, $c$, regulating one output $g$ \cite{twb}. In Fig~\ref{f-oneinput}, we plot the two optimal regulation functions for an activated and repressed gene. These results correspond to two well-defined optima in $I(c;g)$ as a function of the two parameters defining the input/output function, the cooperativity $h$ and the dissociation constant $K_d$. These optimal solutions result from the balance between the two (input and output) components of the noise that limit the information transmission at different values of $c$: the solutions  are a compromise between avoiding readouts at low input concentrations, where input noise is largest (pushing $K_d$ and $h$ to higher values), and being able to distinguish different levels of outputs reliably (pushing $h$ and $K_d$ lower). Because the form of the noise is different for repressed and activated genes, the two optimal input/output relations are not mirror images of each other. However, the capacities of an activated and repressed gene are comparable, with the slight advantage of activated genes over repressed genes increasing as the resources become scarcer (for smaller $C$). 

\begin{figure}
\includegraphics[width =  \linewidth]{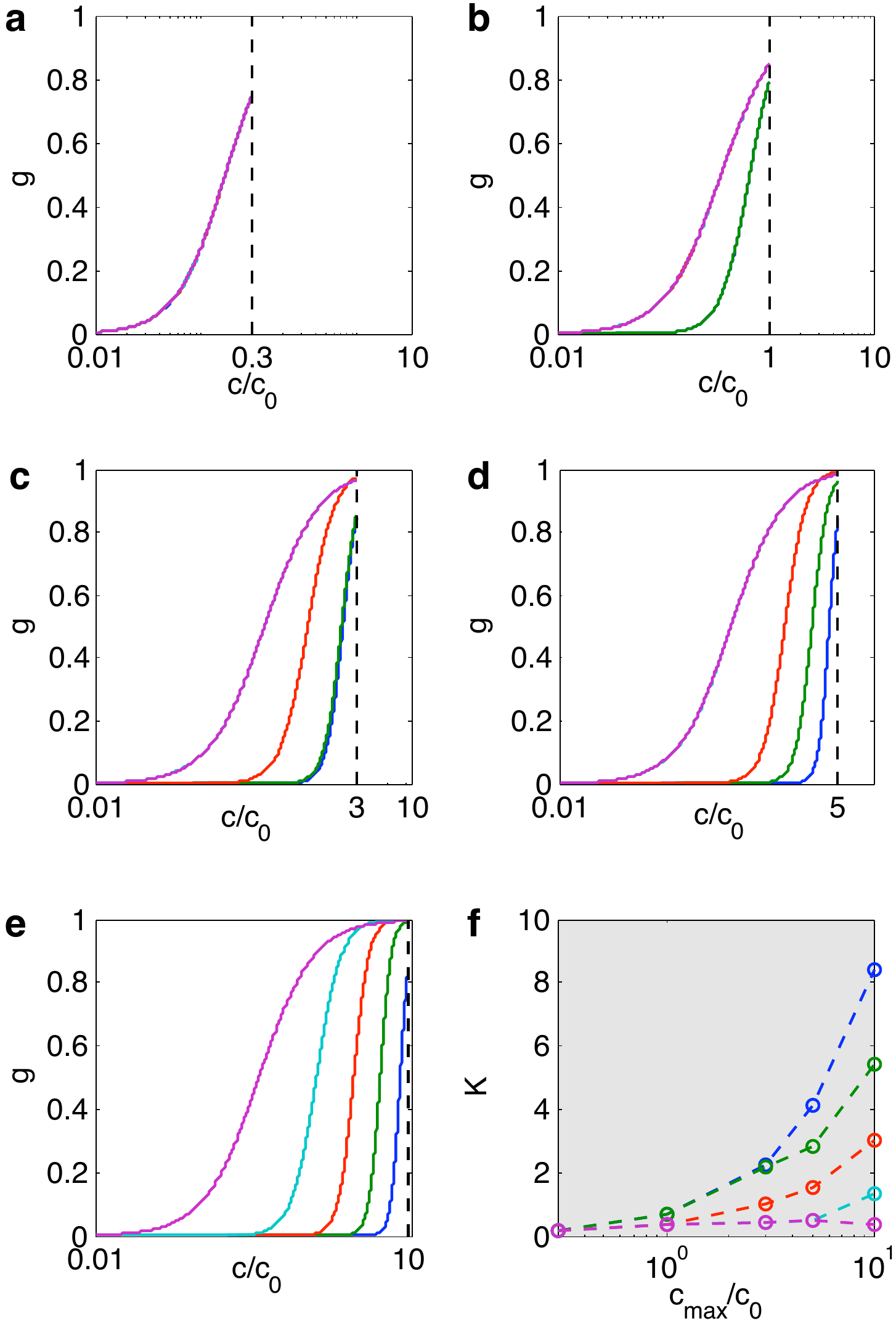}
\caption{The optimal input/output relations for $K=5$ genes, $\{g_1(c),\dots,g_5(c)\}$ (shown in various colors), regulated independently by a common input, $c$. The first 5 panels show optimal solutions depending on the dynamic range of the input, $C$, that is, when $c\in [0,C]$. As $C$ is increased, the totally redundant solution, where $\bar{g}_1(c)=\dots=\bar{g}_5(c)$, slowly becomes non-redundant and transitions into the tiling solution at high $C$, where each $g_i$ independently covers a subrange of concentrations for the input $c$. The last panel shows the optimal values for the dissociation constants, $K_i$, of all 5 genes, as a function of $C=c_{\rm max}/c_0$. }
\label{f-5genes}
\end{figure}

Figure~\ref{f-5genes}  shows the example solutions for $K=5$ noninteracting genes as a function of $C$. We see that there are two regimes: at low $C$, the optimal solutions  for all 5 genes  have exactly the same parameters, and therefore their input/output curves overlap perfectly. Why is this behavior optimal, if at first glance all the genes appear completely redundant? At low $C$, the input noise is dominant, and the best strategy is to have all $K=5$  genes read out the input $c$ and lower the input noise by averaging: using $K$ readouts should lower the effective noise by a factor of $\sqrt{K}$.

At high $C$ another strategy, called the \emph{tiling} solution, becomes optimal: here, each gene $g_i$ changes its expression considerably over some limited range of inputs, and various genes $g_i$ encode various non-overlapping input ranges; in other words, each $g_i$ ``reports'' on its own range of inputs, while the other $g_j$ have either not switched on yet, or are already saturated. We can explore the transition from redundant to tiling solutions in detail, and we can carefully study the scaling of information capacity $I(c;\{g_i\})$ with the number of genes $K$ in each solution \cite{twb}.

Although interesting from a theoretical perspective, the redundant and tiling solutions are not what is actually observed in the real gap-gene network of \emph{Drosophila}. In particular, when $\{g_i\}$ are independent, the only possible input/output relations are sigmoid; there are no stripe-forming solutions, where $g_i$ would turn on at some concentration $c$ and turn off at some higher concentration. Can such solutions emerge if the activating and repressing interactions between the output genes are allowed?

\begin{figure}
\includegraphics[width = 3in]{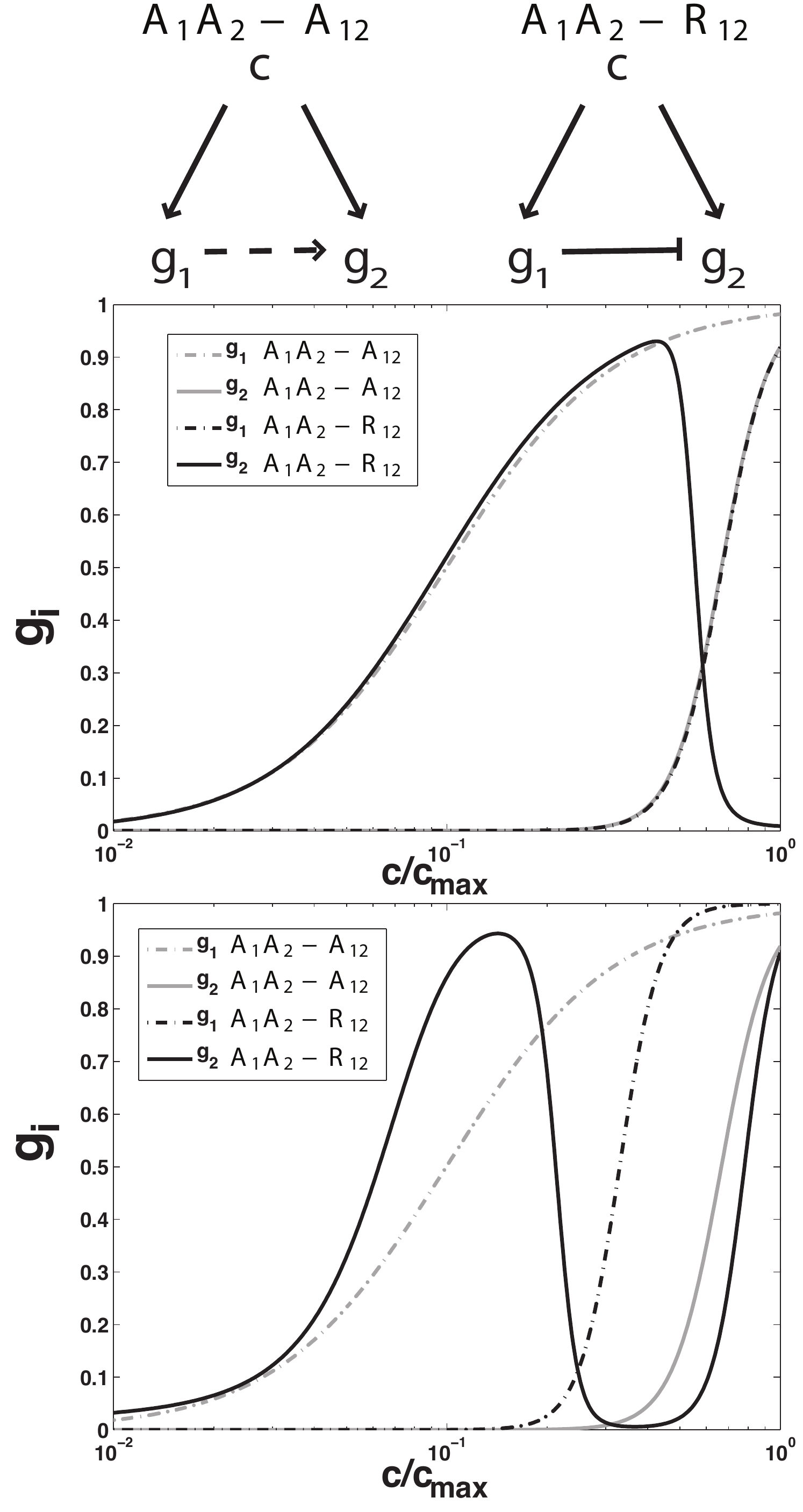}

\caption{The optimal input/output relations for two genes $g_1,g_2$ regulated by a common input $c$, with cross-regulatory feed-forward interactions and Hill model of regulatory functions (top) and the MWC model (bottom). In case the activating arrow is allowed between $g_1$ and $g_2$, the optimal solution (gray lines, $A_1A_2-A_{12}$) is not different from a non-interacting system, where $c$ independently regulates $g_1$ and $g_2$: both the input/output curves as well as the information transmission values are the same. In the case where $c$ activates $g_1$ and $g_2$, but $g_1$ can repress $g_2$, qualitatively new input/output shapes  can be optimal (black lines, $A_1A_2-R_{12}$). Here, the combinatorial regulation of $g_2$ by $g_1$ and $c$ makes the apparent input/output relation $\bar{g}_2(c)$ behave non-monotonically and produce a stripe. }
\label{f-combgenes}
\end{figure}

Indeed we find that this is the case, as shown in Fig~\ref{f-combgenes}. If the interactions between two output genes $\{g_1(c),g_2(c)\}$ are allowed (and optimized over), the information maximizing wiring diagram includes ``lateral repression'' between the two genes that are jointly activated by a common input. This also generates effective input/output curves that are non-monotonic in $c$: $g_2$ as a function of $c$ is seen to exhibit a stripe of activation. Further work has confirmed that such stripe-like patterns optimize information transmission \cite{wtb_pre10}. Interestingly, a similar pattern of interconnections (``lateral inhibition'') is known to occur in neural networks involved in the retinal processing of visual stimuli, and is thought to serve the function of removing redundancy in the neural code due to correlations in the stimulus and receptive field overlap. The function of such connections in genetic regulation is to decrease the redundancy in the outputs as well -- with no interconnections in the tiling solution, when the gene with the highest $K_d$ is saturated and fully active, we \emph{know} that all the other genes are also fully on and saturated: they are therefore providing redundant information. In other words, when there is no interactions, the only patterns of activation\footnote{ These ``patterns'' are defined in a simplified picture when the genes are binary -- 1 if fully active (above threshold) and 0 if inactive (below threshold). In reality the activation functions are real-valued, but conceptually some phenomena are easier to understand if we think of genes being either ``on'' or ``off''.} are $000,001,011,111$ for a case of 3 genes. Patterns such as $010$ or $110$ cannot be accessed if there is no lateral interactions. If they exist, however, these patterns can be generated and they can encode additional useful information about their input $c$, increasing information transmission.

The results we summarized so far are for the case of a functional model of regulation desribed by combinatorial Hill regulation with \emph{AND} logic [Eq~(\ref{andreg})]. As we mentioned in Section \ref{Sec_molmodels}, other phenomelogical models such as the MWC model can be used to described the activation of one gene by many transcription factors. In the lower panel of Fig~\ref{f-combgenes} we plot the optimal network for the same two interacting genes as in the upper panel, but we describe their regulation via a MWC model instead of a combinatorial Hill model. Although the wiring of the optimal networks in the two cases of regulatory models is the same, the input/output relations for the case of MWC regulation exhibit both genes in the ``on'' state for large concentrations of input $c$. In the simplified picture where we view the genes as being ``on'' and ``off'' only, this ``on''/``on'' state affords the MWC model another distinguishable state that encodes the input, and thus results in MWC model achieving a higher information capacity in comparison to the Hill model. The non-interacting solutions for the two regulatory functions are the same.

At this point we would like to stress again what are the assumptions and what are the results of the approach presented here. We assume that {\bf (i)} the information is optimized, {\bf (ii)} the small-noise approximation is applicable, {\bf (iii)} the input/output functions come from a fixed family (of, for instance, Hill or MWC regulatory functions), and {\bf (iv)} the form of the noise is fixed to have an input and an output component, as in Eqs~(\ref{noisefinal},\ref{noisen}) -- the last assumption introduces a single tunable dimensionless parameter, $C=c_{\rm max}/c_0$, on which the optimal solutions depend. What we find from the optimization calculation  is whether a given gene is regulated or not by a given input (an optimized result of $K_i^j=0$ or $n_i^j=0$ means there is no interaction, even if we allowed for one, from gene $i$ to gene $j$), whether the interaction is activating or repressing (sign of $n$), and specifically what is its strength (values of $K$ and $n$). Therefore we learn both the topology of the optimal network and the directions and strengths of the ``arrows'' in its wiring diagram.

Our understanding of information transmission in transcriptional networks is far from complete. Nevertheless, the richness of solutions and network topologies that emerges from a single optimization principle in a one-parameter ($C$) problem is very encouraging, especially since we already observe a  qualitative match to the stripe-like solutions in early \emph{Drosophila} development. Further efforts need to be invested into understanding multi-stability, feedback loops and autoregulation, and in the incorporation of other biologically realistic details. Hopefully, this (or some other) design principle will in the future enable us to understand the wiring of biological networks and derive it from a mathematical measure of their function, rather than reconstructing it back  from painstaking molecular disassembly of the network into its constituent parts.

\subsection{Beyond the small noise approximation}\label{beyondSNA}
The results presented in the previous sections were computed in the small noise approximation, i.e. the assumption that the system is well-described by the set of mean input/output relations along with a (small) Gaussian noise envelope. However, in real networks the small noise approximation might not be applicable for two reasons: first, the noise might be Gaussian in form but not small compared to the mean, and second, the noise might not have a Gaussian distribution. The first possibility was raised already in Refs~\cite{ggpre,ggpnas}, where we showed that in the real bcd/hb system the small-noise result and the exact result are similar, but  the small-noise approximation underestimates the capacity by about $\sim 25\%$. In this section we review more abstract work which analyzes the general properties of information transmission in regulatory elements, without making any assumptions about the form of the noise. 

\begin{figure}
\includegraphics[width =  \linewidth]{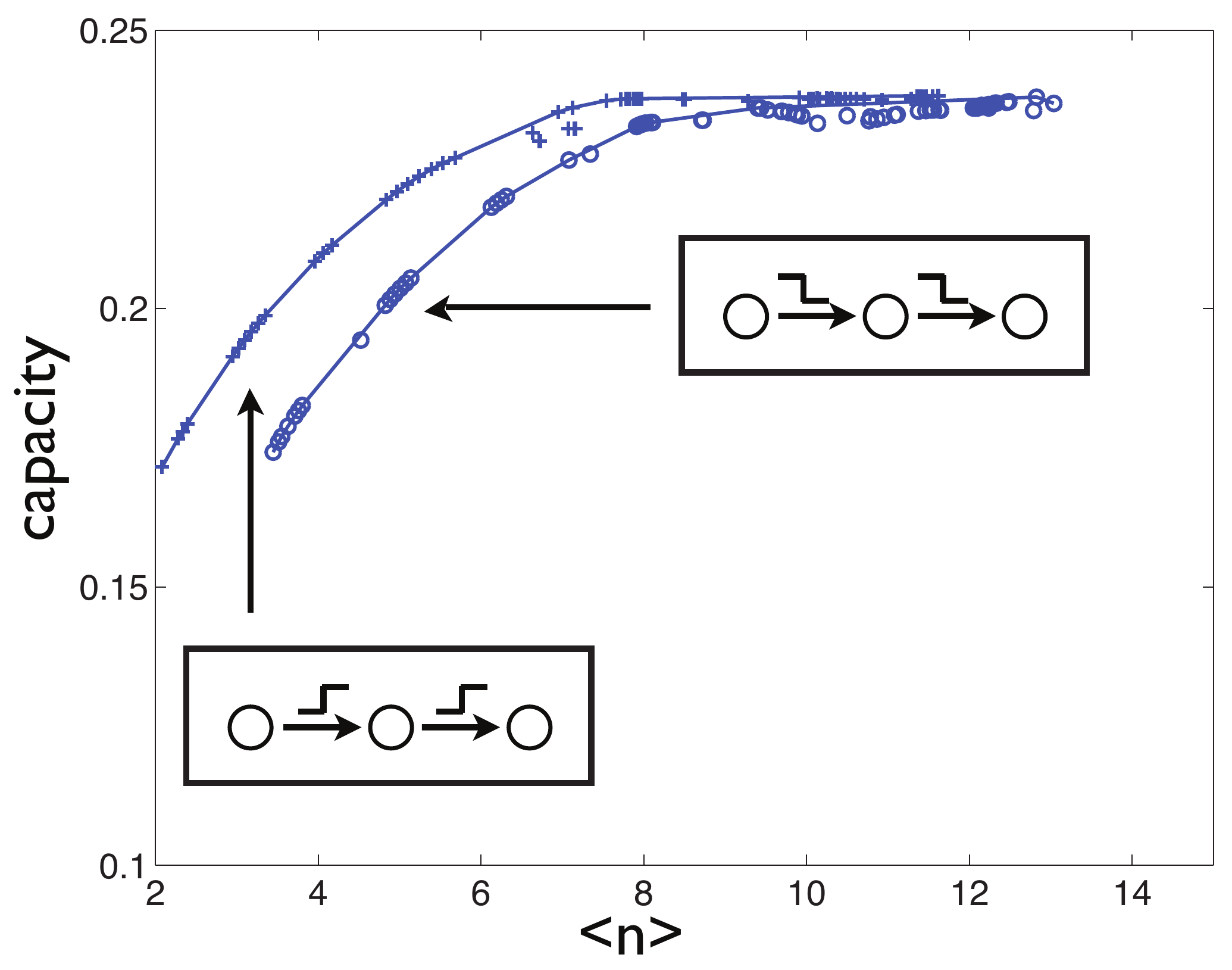}
\caption{The capacity of a cascade of length $L=3$ as a function of the mean number of proteins produced in the cascade, $\langle n\rangle$. A comparison of cascades where in each step the downstream gene is activated (crosses) and repressed (circles). At a fixed (and low) total protein number,  the cascade with positive regulation yields higher capacity than the cascade with negative regulation. These results are derived using the master equation model, with threshold (positive or negative) regulation in each step of the cascade. The input to the cascade is assumed to be Poisson with an optimized mean.}
\label{f-infvscost}
\end{figure}

We start by writing down the full stochastic model for the regulatory circuit of interest using a master equation, which will be a generalization of Eq~(\ref{mastereq}) to more than one gene. The information transmission $I$ between the input and output of a circuit is then computed directly from the definition in Eq~(\ref{mut2}), subject to the  constraint on the mean total number of produced signaling molecules. Specifically, we are maximizing ${\cal L}=I-\lambda \sum_{\ell=1}^L \av{n_{\ell}}/L$, where $L$ is the number of signaling protein species, $n_\ell$ are their counts and $\lambda$ is the Lagrange multiplier enforcing the constraint. An example in Fig~\ref{f-infvscost} shows the capacity results for a two-step regulatory cascade. In general, with this approach the computational difficulty lies in  solving for the steady state probability distribution of a master equation. Since the goal is to optimize the information transmission (which requires many evaluations of the steady state distribution for different choices of parameters and inputs), one must have a fast and accurate method for solving the master equation. For this purpose we derived  the spectral method \cite{wmw, mww}, which is reviewed in detail in Ref~\cite{walczak_chapter}. 

In Section~\ref{ONA} we found that information transmission is increased if  the system is able to access  distinguishable gene expression states. Consequently we wondered if there exists a scenario where an optimal network would transform a unimodal input into a bimodal output. We specifically considered a cascade of length $L$, where at each step the  regulation was taken to be stepwise: there were two protein production rates, one above the threshold, $q_+$, and one below the threshold, $q_-$. We found that for large enough jumps in regulation, $\delta=\abs{q_+-q_-}$, the optimal way to transmit information in a cascade of $L\geq3$ is to generate a bimodal output. For a fixed value of the jump parameter, cascades of repressed genes transmit the same amount of information as activated genes. However, a cascade with repressed genes needs to produce more proteins to achieve the same capacity as a cascade with activated genes (see Fig.~\ref{f-infvscost}). The difference is most significant when we restrict the total mean number of available proteins to be small. In this regime of constrained resources, the master equation approach is especially useful. We also observed how information decreases with every step of the cascade, as expected from the data processing inequality presented in Section~\ref{info}.

\begin{figure}
\includegraphics[width =  \linewidth]{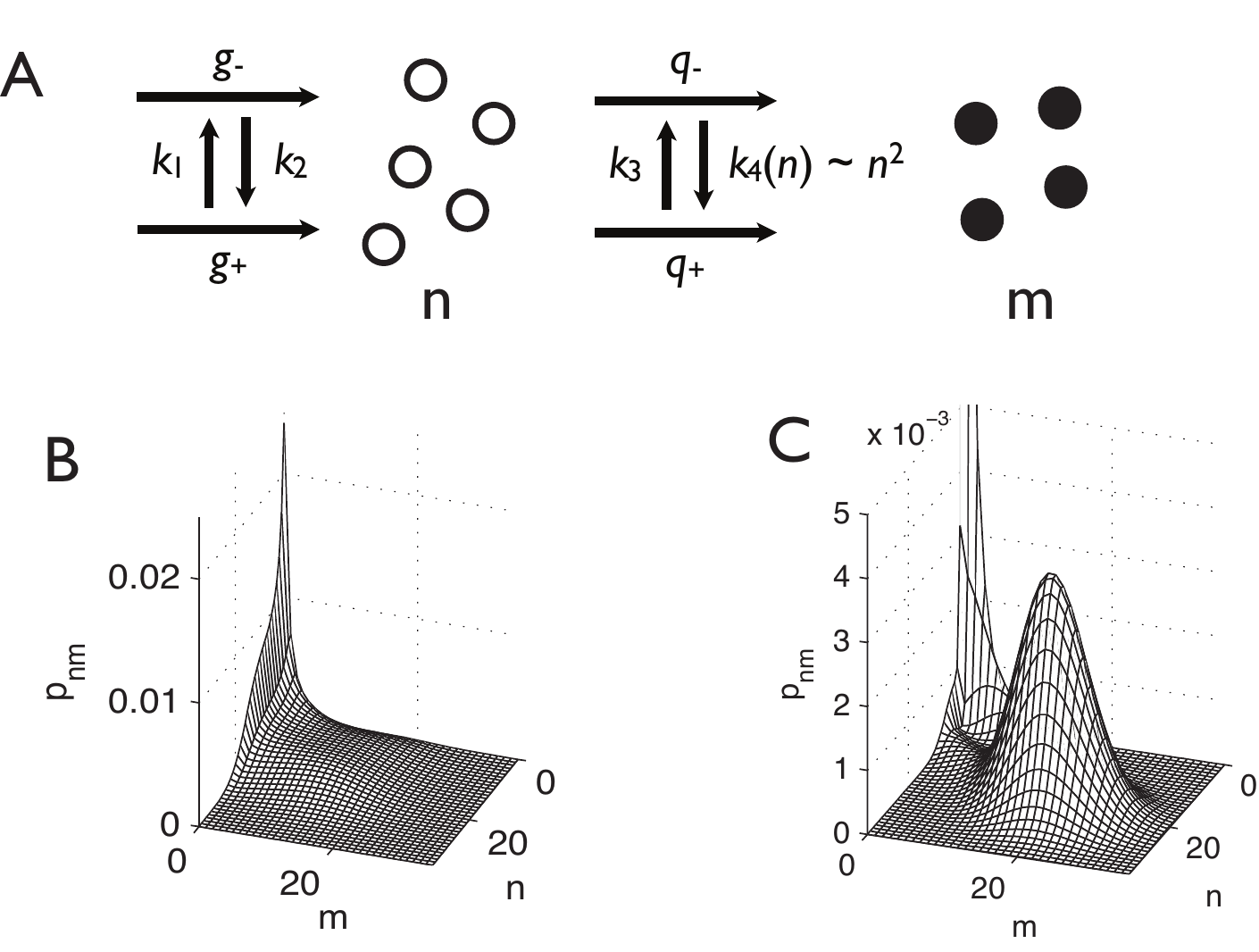}
\caption{ {\bf A)} A regulatory scheme for a two-step cascade, where external input controls the level of gene 1 (with protein count $n$), which in turn controls the production of protein 2 (with protein count $m$). Both genes have two states: an activated and a basal level of expression.  {\bf B)} and {\bf C)} show the two optimal distributions $P_{nm}$ in two regimes of circuit operation: the unimodal distribution in B is optimal for fast switching between the two expression states, and the bimodal distribution in C is optimal for slow switching between the two states. In the limit of slow switching between the two gene expression states the optimal circuit transmits more information between the input and output than in the limit of fast switching. }
\label{f-bimod}
\end{figure}

The issue of how a limit on the number of available signaling molecules (signaling cost) affects the choice of  optimal regulatory functions appeared already in Section~\ref{ONA}  in the form of $C$, the maximal concentration of input molecules. In Ref~\cite{mww} we investigated a detailed model of gene regulation, which explicitly considered two gene expression states, at a basal and enhanced expression level. Surprisingly, we found that when the gene expression state changes on slow timescales, the information transmitted is larger than if the gene expression state is equilibrated (see Fig.~\ref{f-bimod}). This serves as yet another example of how capacity can be increased without increasing the cost, by making a clever use of the regulatory mechanisms  (as was the case with the ``lateral inhibition''). The slow change in the gene expression state generates a  bimodal output distribution, as opposed to a unimodal distribution in the equilibrated case.

\subsection{Beyond the static and steady state assumptions}\label{timedep}
The discussion so far has been limited to the steady state solutions and signals that are static or  that vary in time slower than all other time-scales in the problem. One of the remaining theoretical challenges is to understand information transmission in a fully dynamical, non-linear system. In this case one might focus on various quantities, for instance the instantaneous or mean information rate, or in the total information transmitted as a function of time. In this section we briefly mention recent approaches that explored some of these quantities in certain tractable limits.

As mentioned in the Introduction, a prominent example of time-varying signals occurs in  the oscillatory behavior of TFs involved in the cell cycle. In this system, one can ask about the ability of the genes downstream from the cell cycle oscillator to tell the phase of the signal, which is assumed to change in a sinusoidal  fashion, $f(t)=g+\alpha \sin(\omega t)$. In Ref~\cite{mww10} we considered a two-gene circuit where the expression level of the the first gene is $f(t)$; its products in turn regulate the second (output) gene via a threshold function. The information between the output protein count $m$ and the phase $\phi$ can be calculated knowing the oscillatory steady state solution for the output probability distribution:
\beq
I(\phi; m)=\int_0^{2\pi} d \phi \sum_m P(m|\phi) P(\phi) \log{\frac{P(m|\phi)}{P^0_m}},
\eeq
where $P(\phi)=1/2\pi$ and $P_m^0=\int_0^{2\pi} d \phi P(m|\phi) P(\phi)$. For threshold regulation, the information is optimized for a certain non--zero value of the driving frequency. For infinitely  slow oscillations, the system discriminates between three states in output expression: high, low and intermediate. In the limit of very fast oscillations, all the expression states are averaged together and become indistinguishable, forcing the information content to decrease. In an intermediate regime, two new intermediate states appear in addition to the high and low state.  The output is now able to encode whether the signal is increasing or decreasing. We again find that the ability of the output to discriminate between different states allows for the output protein concentration to carry more information about the phase of the oscillatory signal. We note that in this case the information between the output count $m$ and the phase can be larger than between the upstream protein count and the phase. The data processing inequality does not hold, because the Markovian steady-state assumptions used to derive it are not valid in this case.

Lastly, we turn to an approximate result in the case of a fully time-dependent solution. In Section~\ref{info} we derived the result for a Gaussian channel, where the information capacity was determined by the signal-to-noise ratio. This result can be generalized to a time-dependent stationary system by moving to the Fourier space and observing that the information capacity is now simply an integral of a frequency-dependent signal-to-noise ratio across all frequency channels \cite{cover+thomas_91}. While this result has been in use in engineering and neuroscience for quite some time,   it has been introduced  in the context of gene regulatory networks in a series of pedagogical papers \cite{tostevintenwoldeprl, tostevintenwoldepre, derondetostevintenwoldepre}. We briefly outline the derivation presented in these papers here, while refering the reader to the original manuscripts for detailed discussion and limiting cases.

In general the mutual information between a time dependent input trajectory, $c(t)$, and output trajectory, $g(t)$, is given by a generalization of Eq~(\ref{mut2}):
\begin{eqnarray}
I\left[c(t),g(t)\right]&=&\int {\cal{D}}[c(t)] \int {\cal{D}}[g(t)] P\left(c(t), g(t) \right)\nonumber  \\ 
&&\log_2\frac{P\left(c(t), g(t) \right)}{P\left(c(t))P(g(t) \right)}.\label{infotraj}
\end{eqnarray}
In order to evaluate the integrals one needs to consider all possible paths, which makes the problem extremely hard.  One possible approach is to assume that the input, the output, as well as the (additive) noise in the channel jointly obey Gaussian statistics. Defining the deviations of input and output from their respective means, $\delta c(t)=c(t)-\bar{c}$ and $\delta g(t)=g(t)-\bar{g}$, we can sample the trajectories of the deviations at $N$ successive, evenly spaced points in time, $\{t-(N-1)\Delta,t-(N-2)\Delta, \dots,t-\Delta, t\}$, pack them into vectors $\delta \mathbf{c}(t)$ and $\delta \mathbf{g}(t)$, and write down the joint probability distribution for these vectors:
\begin{eqnarray}
P\left( \delta \mathbf{c}(t), \delta \mathbf{g}(t)\right)&=&(2\pi)^{-N} \abs{\mathbf{C}}^{-1/2} \times\label{multgaussian} \\
&\times&\exp\left({-\frac{1}{2} \left[ \begin{array}{c}
 \delta \mathbf{c}(t) \\
 \delta \mathbf{g}(t)\\
\end{array} \right]^T \mathbf{C}^{-1} \left[ \begin{array}{c}
 \delta \mathbf{c}(t) \\
 \delta \mathbf{g}(t)\\
\end{array} \right]}\right),\nonumber
\end{eqnarray}
where the time dependent covariance matrix has the form:
\beq
\mathbf{C}(t,t')=\left( \begin{array}{cc}
\mathbf{C}_{cc} &\mathbf{C}_{cg}  \\
\mathbf{C}_{cg} &\mathbf{C}_{gg}  \\ \end{array} \right).
\eeq
The elements of the covariance matrix are the correlations between having, e.g., a concentration of input $\delta c(t)$ at time $t$ and a concentration of output $\delta g(t')$ and time $t'$, $C_{cg}(t, t')=\av{\delta c(t) \delta g(t')}$. Each submatrix $\mathbf{C}_{\mu\nu}$ of $\mathbf{C}$ with $\mu,\nu=\{c,g\}$  is of dimension $N\times N$.

We note that the assumption of joint gaussianity of inputs and outputs is much stronger than the small noise approximation we used in the steady state analysis of Section~\ref{SNA}. There we only assumed that the noise profile $\sigma_g(c)$ is locally a Gaussian at every $c$ around the mean input-output relation $\bar{g}(c)$ which itself could be arbitrarily nonlinear. Here, we are making a stronger approximation that across the whole dynamic range of inputs and outputs and across time the distribution is jointly Gaussian; as a result  we gain the ability to consider time-dependent signals. Since we are dealing with Gaussian distributions, the entropy is proportional to the logarithm of the variance, $S=\log{\sqrt{(2 \pi e)^N \abs{\mathbf{C}}}}$, as we showed following Eq~(\ref{Gaussian_c}). Plugging in Eq~(\ref{multgaussian}) into Eq~(\ref{infotraj}) one obtains:
\begin{eqnarray}
I(\mathbf{c};\mathbf{g})&=&S[P(\delta\mathbf{c})]+S[P(\delta\mathbf{g})]-S[P(\delta\mathbf{c},\delta\mathbf{g})]\label{inf_time} \\
&=&\frac{1}{2}\log_2{\frac{\abs{\mathbf{C}_{cc}}\abs{\mathbf{C}_{gg}}}{\abs{\mathbf{C}}}}.\nonumber
\end{eqnarray}
When the conditional probability $P(g|c)$ is not Gaussian, the Gaussian channel approximation remains a lower bound on the amount of information that can be transmitted between the input and the output. This calculation therefore remains a very useful first step to gaining intuition about the properties of any system.

In the context of time-varying signals, the amount of information transmitted is proportional to the duration of signal transmission. The quantity we really are interested in is the average information rate, which we define as:
\beq
R\left(\mathbf{c}; \mathbf{g}\right)=\lim_{T\rightarrow \infty} \frac{I\left(\mathbf{c};\mathbf{g}\right)}{T}.\label{rateinfodef}
\eeq
The information rate has units of [bits/sec]. Since in gene regulation we most often deal with continuous signals, such as concentrations which fluctuate in time, it is convenient to continue the analysis in the Fourier domain. We are interested in a time-averaged information rate, and in the $T\rightarrow\infty$ limit we  can restrict our analysis to stationary signals $C(t,t')=C(t-t')$.  We can thus rewrite the covariances in terms of their Fourier transforms, the power spectra, e.g.
\beq
S_{cg}(\omega)=\int_{-\infty}^{\infty} d(t-t')\;C_{cg}(t-t') e^{i\omega (t-t')}.
\eeq 
Next, we rewrite Eq~(\ref{inf_time}) in Fourier basis and calculate the information rate from Eq~(\ref{rateinfodef})\footnote{We note that the Fourier transform of $\log C$ for the Gaussian input-output correlation matrix $C$ is $\int d \omega \log[S_{cc}(\omega)S_{gg}(\omega)-\abs{S_{cg}(\omega)}^2]$.}:
\beq
R(\mathbf{c};\mathbf{g})=-\frac{1}{4 \pi} \int_{-\infty}^{\infty} d \omega \log_2 \left[ 1-\frac{\abs{S_{cg}(\omega)}^2}{\abs{S_{cc}(\omega)}\abs{S_{gg}(\omega)}}\right].
\eeq
The power spectrum of the output can be written in terms of the power spectra of the noise, $N(\omega)$, and the transmitted input, $\Sigma(\omega)=\abs{P_{cg}(\omega)}^2/P_{cc}(\omega)$:
\beq
S_{gg}(\omega)=\Sigma(\omega)+N(\omega).\label{outputnoise}
\eeq
We can now rewrite the rate of information transmission in terms of the signal--to--noise ratio, $\Sigma(\omega)/N(\omega)$:
\beq
R(\mathbf{c};\mathbf{g})=\frac{1}{4 \pi} \int_{-\infty}^{\infty} d \omega \log_2 \left[ 1+ \frac{\Sigma(\omega)}{N(\omega)}\right].
\label{ratefinal}
\eeq
This result is a generalization of the Gaussian channel to time dependent stationary signals. If there were no frequency dependence in the signal-to-noise ratio, we would recover the result of Eq~(\ref{gaussian_bound}). In case of stationarity, Fourier components are statistically independent and the total information rate is the summation across all frequency bands.  Using that fact, we can motivate the time dependent result by taking the total information transmitted to be a sum of all the Fourier components, $I_n$ transmitted independently \cite{spikes}, 
\beq
I=\sum_n I_n= \frac{1}{2} \sum_n \log_2 \left[1+\frac{\Sigma(\omega_n)}{N(\omega_n)}\right].
\eeq
Taking the limit of continuous frequencies we arrive at the known form for the rate of information transmission in Eq~(\ref{ratefinal}). In this approximation, a signal at a given frequency will only trigger a response at that same frequency, i.e. there is no ``frequency mixing.'' This result allows us to optimally choose the signal power spectrum so that we maximize the rate of information transmission through the system with a given noise spectrum $N(\omega)$. The answer [see Ref.~\cite{spikes} for a derivation] is for the signal to be complementary the noise, that is, chosen such that $\Sigma(\omega)+N(\omega)=\mathrm{const}$ (``waterfilling''). For a finite signal power, this will make the combined spectrum in an optimal case look flat and unstructured. 

This framework has, up to now, been completely general. The application to biological signaling consists of computing the  covariance structure of inputs and the outputs that enters Eq~(\ref{inf_time}) for the specific system under study. One approach, presented in the work of Tostevin and ten Wolde, is to calculate them from the linear noise approximation \cite{tostevintenwoldeprl}. In the linear noise approximation the dynamical equations for the system  are linearized around their operating point, to yield a linear system with a Gaussian additive noise exposed to Gaussian inputs (by assumption), so that the statistics of inputs and outputs will be jointly Gaussian. One can then calculate all the covariances in the system, and finally compute the information rate, as described. For a discussion of the validity of the linear noise approximation see Ref~\cite{walczak_chapter}. 

An important result demonstrated by Tostevin and ten Wolde is that a gene regulatory circuit for which the instantaneous information is zero can have a large information rate for the input/output trajectories, and vice versa. An example of such a system is the irreversible conversion of one molecule into another. Therefore, a  gene circuit may not be transmitting information is its stationary state, yet it could transmit information in an oscillatory state.  The authors also discuss that when gain-to-noise ratio depends on the statistics of the input, the optimal input power spectrum no longer needs to obey the simple waterfilling rule. Later, de Ronde and co-workers \cite{derondetostevintenwoldepre} have studied systematically information transmission in short regulatory motifs with and without feedback. They explore the following interesting signaling cascades: with and without positive or negative feedback, and adding feedback such that it either acts at from the output node or upstream in the circuit. They showed that  negative feedback from the output onto intermediate stages of the cascade is not a good strategy for transmitting information. Similarly, positive auto-regulation of the output node does not increase information transmission, while positive auto-regulation of an intermediary node does. The effect of feedback between intermediate nodes depends on the type of feedback: negative feedback increases transmission fidelity at high frequencies (but across the whole bandwidth, the gain-to-noise ratio decreases overall), while positive feedback increases gain-to-noise at low frequencies.
This detailed study lead the authors to claim that, in general, feedback, including auto-regulation, can increase the circuits ability to transmit information between the input and output, but only if these forms of regulation occur upstream of the dominant source of noise.

Gene regulation is a highly nonlinear process for which the linear noise approximation can fail. A simple signature that invalidates the linear noise approximation is a bimodal distribution of either inputs or outputs (as in the case of bcd/hb system), where the mean will be a poor representation of either of the two states, and the variance will be badly approximated from the linear expansion around the mean. On the other hand, there are signaling networks that might operate close to the linearized regime, e.g. the chemotaxis network of \emph{Escherichia coli}.  As noted in Ref~\cite{tostevintenwoldeprl}, the linear noise approximation is a natural choice for information rate calculation in the Gaussian approximation, because linearization of the dynamical system also decouples frequency components. We refer the reader to the original work in Refs~\cite{tostevintenwoldeprl, tostevintenwoldepre, derondetostevintenwoldepre} for derivation of covariance matrices and information rates for common motifs in regulatory networks. In gene regulatory networks, covariances can have complex forms and the output power spectrum can depend on both the statistics of the input and the noise \cite{sorin}.

\section{Related work}
We focussed on one specific approach to information processing by gene networks, where we optimize the form of the regulatory function to maximize the information between the input and output. 
This approach has been enormously successful in sensory neuroscience, where it is known as the ``efficient coding'' principle \cite{barlow}. Assuming that the statistics of the input  signal are fixed by the environment, the neural processing mechanisms have evolved to transmit as much of the input information as possible through noisy neuronal links of limited bandwidth. This has led to a number of predictions regarding the structure of receptive fields \cite{attick}, design of the retinal mosaic \cite{devries, borghuis,liu,onoff}, and properties color vision \cite{osorio,conearray}. The same principle has also been invoked to predict that neurons should dynamically adapt to the modulations in stimulus statistics. Impressively, experiments have confirmed that the neurons in fly vision really do scale their input/output relations so as to match their dynamic range to the variance of the stimulus and thus increase information flow \cite{brenner,fairhall}. Recent work has also examined the nature of optimal population coding in neural networks, exploring the tradeoffs between fighting the noise through positive couplings and reducing redundancy through lateral inhibition; it is interesting to note that in some parameter regimes optimal codes again turn out to be locally stable and distinguishable states of the output \cite{tkacikcodes}. Advances have also been made in constraining information encoding by network elements with experimental measurements \cite{globerson}. These parallels between genetic regulation and neuroscience certainly motivate us in thinking that the same set of basic principles might underlie efficient biological information processing.

Other applications of information theory to cell regulation have been developed, which consider bounds on information transmission in biological systems, such as finding the minimum rate at which information must be transmitted in the system to ensure the readout of the signal remains within a fixed value of the signal -- these bounds have much to do with the approach that views cells and organisms as trying to ``decoding'' noisy environmental signals  and making optimal decisions based on these data \cite{libby}. Information theory has been used to discuss chemotaxis \cite{shraiman, iglesias}, i.e. navigation on the basis of noisy inputs.

A recent paper has also raised an interesting topic of learning about biological systems from the way they systematically deviate from the optimality predictions \cite{polavieja}.

For a (nonexhaustive) list of other topics where information theory has also been applied in biology, we note its use in analyzing evolution of organisms in unknown environments \cite{KusselLeibler, sambill, olivier}, or considering the capacity of genomes \cite{Mirny}. It has been discussed more generally in the context of evolution \cite{maynardsmith}.

\section{Discussion}\label{Outro}
Biology presents an interesting challenge to physicists: many symmetries and simplifications applicable in ordered (but non--animate) systems are absent in biology, and this complexity of life can be intimidating. On the other hand, biological systems have evolved for function, and as we make progress in formalizing this notion mathematically, we hope to gain new insights and predictive power. 

In this review we attempted to summarize some of the progress made over the last few years in using information transmission as a possible measure of function for gene regulatory networks. Our goal was to show that this is a powerful approach which allows one to calculate properties of gene circuits that can be directly compared to measurements. One of the interesting aspects we tried to illustrate in this review is how microscopic features of gene regulation, i.e. the nature of computations / signal integration at the promoter, and the form of the noise in gene regulation, influence the ability of the network to transmit information and affect the pattern of optimal solutions (e.g. stripes of expression in the gap gene network). To physicists this is an interesting lesson, in that some features at the macroscopic level can be sensitive to certain (hopefully not all!) microscopic details of regulation. Nevertheless, the calculations and ideas presented in this review show that attempts at the interface of physics and biology aimed at understanding how physical constraints  shape circuit structure and function are proving fruitful.

 We also described recent experiments that discuss a specific gap gene circuit, active during early development of the fly embryo, which appears to function close to the limits imposed by noise in gene expression. We emphasize again that not all gene regulatory networks are likely to be optimized for information transmission, but in the case of early development, it does seem possible that the formal notion of information reflects faithfully the developmental ``positional information'' that enables the organism to build up complex structures.

Despite progress in understanding information transmission in gene regulation, a lot of work still remains to be done. The biggest formal challenge is to construct a general framework for computing information transmission in time-dependent non-linear networks. The second important challenge is to understand how information transmission functions in spatially resolved systems where constituent chemicals are not well-mixed, and transport phenomena play an important role. Third, as a challenge to both theory and experiment, we are looking for a complete derivation of an optimal information transmission network that includes all relevant regulatory effects, and compare it to both the experimentally determined network topology \emph{and} the experimentally measured information rates. Lastly, we would like to encourage further work that tries to link information transmission to other measures of network function, both as a numerical optimization problem and in models of evolution under the assumed network function. It seems likely -- especially due to the assumption-free nature of information measures -- that different measures of function could produce consistent results.

\section*{Acknowledgements}
We would like to thank William Bialek, Andrew Mugler and Chris Wiggins for major contributions to the work presented in this review. We thank Curt Callan and Thomas Gregor for their contributions. We also thank the Princeton biophysics community, especially Justin Kinney, Pankaj Mehta and Thierry Mora, for fruitful discussions.

\end{document}